%% file: AA8113.tex
\documentclass{aa}
\usepackage{graphicx}
\usepackage{lscape}
\usepackage{longtable}
\usepackage{txfonts}
\begin{document}
   \title{Old open clusters in the outer Galactic disk
\thanks{Based on observations with
  the ESO VLT at the Paranal Observatory, under the program 076.B-0263
},\thanks{Table 3 is only available in electronic form at {\tt http://www.aanda.org}}}

   \author{G. Carraro
          \inst{1,2}
          \and
          D. Geisler
          \inst{3}
          \and
          S. Villanova
          \inst{1}
          \and
          P. M. Frinchaboy
          \inst{4}
          \and
          S. R. Majewski
                 \inst{5}
          }

   \offprints{G. Carraro}

   \institute{Dipartimento di Astronomia, Universit\`a di Padova,
Vicolo Osservatorio 2, I-35122, Padova, Italy\\
              \email{giovanni.carraro,sandro.villanova@unipd.it}
         \and
ESO, Alonso de Cordova 3107 Vitacura, Santiago de Chile, Chile\\
              \email{gcarraro@eso.org}
         \and
             Universidad de Concepci\'on, Departamento de Fisica,
Casilla 160-C, Concepci\'on, Chile\\
             \email{dgeisler@astro-udec.cl}
          \and
       NSF Astronomy and Astrophysics Postdoctoral Fellow\thanks{Any opinions, 
findings, and conclusions or recommendations 
expressed in this material are those of the 
author(s) and do not necessarily reflect the views of the National Science 
Foundation.}\\
       University of Wisconsin-Madison, Department of Astronomy
       475 N. Charter Street, Madison, WI 53706\\
             \email{pmf@astro.wisc.edu}
           \and
           Department of Astronomy, University of Virginia, P.O. Box
	   400325, Charlottesville, VA 22903-4325\\
             \email{srm4n@virginia.edu}
             }

   \date{Received xxxxx; accepted xxxx}

% \abstract{}{}{}{}{}
% 5 {} token are mandatory

  \abstract
   {The outer parts of the Milky Way disk are believed to be one of the main arenas
  where the accretion of external material in the form of dwarf
  galaxies and subsequent formation of
streams is taking place. The Monoceros stream and
  the Canis Major and Argo over-densities are notorious examples.
  Understanding whether what we detect is the signature of accretion
  or, more conservatively, simply the intrinsic nature of the disk,
represents one of the
  major goals of modern Galactic astronomy.}
   {We try to shed more light on the properties of the outer disk by
  exploring the properties of distant anti-center old open clusters. We want to verify
 whether distant clusters follow the chemical and dynamical
  behavior of the solar vicinity disk, or whether their properties
  can be better explained in terms of an extra-galactic population.}
   {VLT high resolution spectra have been acquired for five distant open
  clusters: Ruprecht~4, Ruprecht~7, Berkeley~25, Berkeley~73 and
  Berkeley~75. We derive accurate radial velocities to distinguish  field
  interlopers and cluster members.
  For the latter we perform a detailed abundance analysis and derive the
  iron abundance [Fe/H] and the abundance ratios of several $\alpha$ elements.}
   {Our analysis confirms previous indications that the radial
  abundance gradient in the outer Galactic disk does not follow the
  expectations extrapolated from
the solar vicinity, but exhibits a shallower slope.
  By combining the metallicity of the
  five program clusters with eight more clusters for which high resolution
  spectroscopy is available, we find that the mean metallicity in the
  outer disk between 12 and 21 kpc from the Galactic center is [Fe/H]
   $\approx
  -0.35$, with only marginal indications for a radial variation.
In addition, all the program clusters exhibit solar scaled   or
  slightly enhanced $\alpha$ elements, similar to open clusters in the solar
  vicinity and thin disk stars.}
   {We investigate whether 
   this outer disk cluster sample might belong to an
  extra-galactic population, like the Monoceros ring.
  However,  close scrutiny of their properties -
  location, kinematics and chemistry - does not convincingly  favor this hypothesis.
  On the contrary, they appear more likely genuine Galactic disk
  clusters. We finally stress the importance to obtain 
  proper motion measurements for these clusters
to constrain their orbits.}

   \keywords{Milky Way : general -Open clusters and associations:
                general - Galactic disk: chemical evolution
               }

   \maketitle

\section{Introduction}
While there are much data on chemical abundances in the solar
vicinity both for disk stars (Ram\'irez et al. 2007, Bensby et al.2004)
and for star clusters (Friel et al. 2002),
our knowledge of the outer Galactic disk is still very
poor. This is particularly true for Galactic open
clusters, which are known to extend  to very far in the disk
periphery, but whose data in many cases basically consist 
of shallow photometry which 
only allows us to derive rough estimates of age and distance
(e.g., Phelps et al. 1994; Frinchaboy \& Phelps 2002; Carraro et al. 2006).

In the last few years the situation has improved thanks to
a renewal of interest in the outer Galactic disk which,
according to several studies, may have been formed though
several mergers of small galaxies. In this context, a
structure like the Monoceros ring (hereinafter MRi, Newberg et al. 2002, Ibata et
al. 2003, Crane et al. 2003, Rocha-Pinto et al. 2003)
would be the best signature of such accretions.
Distant
star clusters --- both open and globular --- may trace this structure
(Frinchaboy et al. 2004, 2006), date its formation time, and probe
its chemical evolution,  but the lack of precise
metallicity, distance and age data for a sufficient number of potentially associated
clusters
makes their connection to such over-densities still vague.\\
Recently, good abundance data have started to be acquired
for a handful of distant clusters (Carraro et al. 2004,  Yong
et al. 2005; Villanova et al. 2005, Frinchaboy et al. 2007).
All these studies seem to indicate
that the outer disk does not follow closely the chemical pattern
one would expect from an extrapolation of the solar neighborhood data.
For instance, the radial [Fe/H] gradient, instead
of relatively steeply declining as in the solar vicinity, deviates at
R$_{GC} > 10-12$ kpc and stays almost flat toward
the Galactic anti-center, while the $\alpha$  elements are moderately enhanced
with respect to the Sun.
The metallicity flattening by the way was suggested as early as the study of Twarog et al. (1997), but
with a  largely inhomogeneous data-set .

These trends are seen in other tracers,
like field giants and Cepheids (Carney et al. 2005, Yong et al. 2006).
But while these chemical characteristics may cast new
light on the formation of the outer Galactic disk, in fact the number of
anti-center tracers with good abundance data, particularly
distant star clusters, remains too small to attempt reliable
speculations,
despite their obvious statistical advantage for deriving ages, kinematics,
abundances, and so forth.

Meanwhile, current models of the chemical evolution of the Galactic disk
(e.g., Cescutti et al. 2006, and references therein) have started to predict
the radial abundance gradients for several elements in the outer
regions of the Milky Way.
While these models employ updated prescriptions
for all of the basic ingredients (reactions rates, Initial Mass
Function and so forth ) of the calculation
--- and in this respect
are very sophisticated and have high predictive power --- again  the lack of
sufficient data for elemental abundances in the outer disk prevents
careful comparisons with observations and  limits the applicability of the models.
\noindent
In an attempt to improve this situation, we present in this paper
the results of a spectroscopic campaign conducted with the Very Large Telescope (VLT) of five
previously unstudied,
old  and distant open clusters toward the Galactic anti-center.  These particular clusters were
targeted to
help clarify their status as genuine disk clusters or as possible members of
the MRi. At the same time, our data are intended to provide a better set
of observational templates for chemical evolution models.

\section{Selection of clusters}
The clusters under investigation are listed in Table~1, together with
their Galactic coordinates and preliminary estimates of distance and age as
reported in Carraro et al (2005a,b).
In those papers, BVI photometry has been obtained with the purpose of searching
for potentially old and distant clusters and providing targets for later
spectroscopic follow-up.\\
Together with these five clusters we have also observed Tombaugh~2,
which we report on elsewhere (Frinchaboy et al. 2007).
We elected to study these targets in detail because we wanted
to probe a region of the Milky Way --- the Third Galactic Quadrant ---
where the signatures of ongoing accretions (the putative Canis Major galaxy, hereafter ``CMa",
and the MRi) have been repeatedly pointed out, but where
other explanations are also possible.
Open clusters
can help us to shed more light on the
structure and origin of the outer disk
because of the ability to derive accurate determinations of age, distance, radial velocity
 metallicity and detailed abundances for these systems.
Establishing any relationships between position and kinematics and/or
age and metallicity is the first step not only to better describing the
chemical and dynamical evolution of the Galactic disk periphery, but also
to recognize possible structures
not associated with the Galactic disk (Frinchaboy et al. 2004).\\
For our study we wanted a sample of clusters covering a wide baseline in
Galactocentric distance, older than the Hyades, located toward
the CMa and MRi over-densities and below the plane to further investigate
the disk warp (Momany et al. 2006).

All five of the Table 1 clusters are projected toward the Monoceros/Canis
Major constellations in the southern Galactic hemisphere.
Indeed, they are all projected onto or very near the Canis Majoris overdensity
itself (see Moitinho et al. 2006). They also all lie significantly below the
nominal Galactic plane.
According to the preliminary analysis by
Carraro et al. (2005a,b), they presently all lie beyond 12 kpc from the Galactic
center and are likely older than the Hyades (see Table~1).

To our knowledge, no previous estimates of radial velocity or  metallicity are
available for any of these clusters.

\begin{table}
\caption{Clusters sample}
\fontsize{8} {10pt}\selectfont
\begin{tabular}{cccccc}
\hline
\multicolumn{1}{c} {$Name$} &
\multicolumn{1}{c} {$l$} &
\multicolumn{1}{c} {$b$} &
\multicolumn{1}{c} {$d_{\odot}$} &
\multicolumn{1}{c} {$age$} &
\multicolumn{1}{c} {$Ref$} \\
\hline
 & deg & deg  &  kpc & Gyr &  \\
\hline
Berkeley~75& 234.30 & -11.12 &  9.8 & 3.0 & Carraro et al. (2005a)\\
Berkeley~25& 226.60 &  -9.69 & 11.3 & 4.0 & Carraro et al. (2005a)\\
Ruprecht~7 & 225.44 &  -4.58 &  6.5 & 0.8 & Carraro et al. (2005b)\\
Ruprecht~4 & 222.04 &  -5.31 &  4.9 & 0.8 & Carraro et al. (2005b) \\
Berkeley~73& 215.28 &  -9.42 &  9.7 & 1.5 & Carraro et al. (2005a) \\
\hline
\end{tabular}
\end{table}

\begin{table*}
 \fontsize{8} {10pt}\selectfont
 \tabcolsep 0.5truecm
 \caption{Radial velocities plus photometry of the program stars.ID is identification
 according to Carraro et al. 2005a,b numbering. The last column
 indicates whether a star is considered member (M) or not (NM)
 according to the analysis in Section.~5}
 \begin{tabular}{cccccccccc}
 \multicolumn{1}{c}{Cluster}         &
 \multicolumn{1}{c}{ID}         &
 \multicolumn{1}{c}{$\alpha (2000.0)$} &
 \multicolumn{1}{c}{$\delta(2000.0)$}        &
 \multicolumn{1}{c}{$B$}       &
 \multicolumn{1}{c}{$V$}       &
 \multicolumn{1}{c}{$I$}       &
 \multicolumn{1}{c}{$RV_H$}      &
 \multicolumn{1}{c}{$n$}       &
 \multicolumn{1}{c}{$Membership$}\\
  \hline
 & & hh:mm:sec & $o$:$\prime$:$\prime\prime$& & & &  [km/sec] &&\\
\hline
Berkeley~75 &9  & 06:49:07.09 & -23:59:44.94 & 15.78 & 14.89 & 13.71 & 73.10$\pm$0.50 & 2&NM\\
            &22 & 06:48:55.85 & -24:00:07.16 & 17.08 & 16.13 & 14.99 & 94.60$\pm$0.35 & 2&M\\
Berkeley~25 &10 & 06:41:20.40 & -16:28:23.63 & 17.45 & 16.13 & 14.56 &134.10$\pm$0.20 & 2&NM\\
            &12 & 06:41:16.60 & -16:28:16.13 & 17.59 & 16.32 & 14.78 &135.60$\pm$0.20 & 2&M\\
            &13 & 06:41:13.63 & -16:29:17.27 & 17.36 & 16.33 & 14.97 &133.30$\pm$0.20 & 2&M\\
            &21 & 06:41:08.37 & -16:29:20.13 & 17.59 & 16.50 & 15.08 &134.00$\pm$0.50 & 2&M\\
            &23 & 06:41:16.76 & -16:29:21.34 & 17.91 & 16.72 & 15.27 &131.70$\pm$0.50 & 2&M\\
Ruprecht~7  &2  & 06:57:50.14 & -13:13:39.93 & 15.47 & 14.07 & 12.71 & 77.12$\pm$0.20 & 3&M\\
            &4  & 06:57:43.08 & -13:12:17.41 & 16.02 & 14.58 & 12.71 & 77.10$\pm$0.25 & 3&M\\
            &5  & 06:57:54.56 & -13:13:59.18 & 16.17 & 14.78 & 12.95 & 77.12$\pm$0.20 & 3&M\\
            &6  & 06:57:53.66 & -13:12:57.83 & 16.24 & 14.79 & 12.91 & 77.61$\pm$0.30 & 3&M\\
            &7  & 06:57:52.85 & -13:12:55.08 & 16.41 & 15.07 & 12.25 & 74.26$\pm$0.90 & 3&M\\
Ruprecht~4  &3  & 06:48:51.69 & -10:30:20.49 & 15.46 & 14.24 & 12.69 & 48.40$\pm$1.00 & 1&M\\
            &4  & 06:49:00.80 & -10:31:57.81 & 15.60 & 14.30 & 12.68 & 45.85$\pm$0.55 & 2&M\\
            &8  & 06:48:55.37 & -10:31:32.39 & 15.94 & 14.77 & 13.26 & 48.40$\pm$1.00 & 1&M\\
            &18 & 06:48:54.51 & -10:33:21.32 & 16.62 & 15.44 & 13.90 & 62.20$\pm$1.00 & 1&NM\\
            &29 & 06:48:57.87 & -10:30:27.09 & 17.29 & 15.95 & 14.25 & 62.20$\pm$1.00 & 1&NM\\
Berkeley~73 &12 & 06:21:54.89 & -06:17:51.91 & 16.30 & 15:30 & 13.93 & 89.10$\pm$0.20 & 2&NM\\
            &13 & 06:21:55.36 & -06:20:10.04 & 16.10 & 15.36 & 14.30 & 75.35$\pm$0.25 & 2&NM\\
            &15 & 06:22:10.44 & -06:19:30.19 & 16.51 & 15.57 & 14.30 & 73.30$\pm$0.30 & 2&NM\\
            &16 & 06:22:01.13 & -06:19:17.14 & 16.86 & 15.84 & 14.48 & 95.30$\pm$0.20 & 2&M\\
            &18 & 06:22:01.99 & -06:19:21.89 & 17.10 & 15.99 & 14.58 & 96.10$\pm$0.20 & 2&M\\
            &19 & 06:21:57.24 & -06:17:15.55 & 16.74 & 16.03 & 14.95 & 92.85$\pm$0.25 & 2&NM\\
 \hline
 \end{tabular}
\end{table*}

    \begin{figure}
   \centering
   \includegraphics[width=\columnwidth]{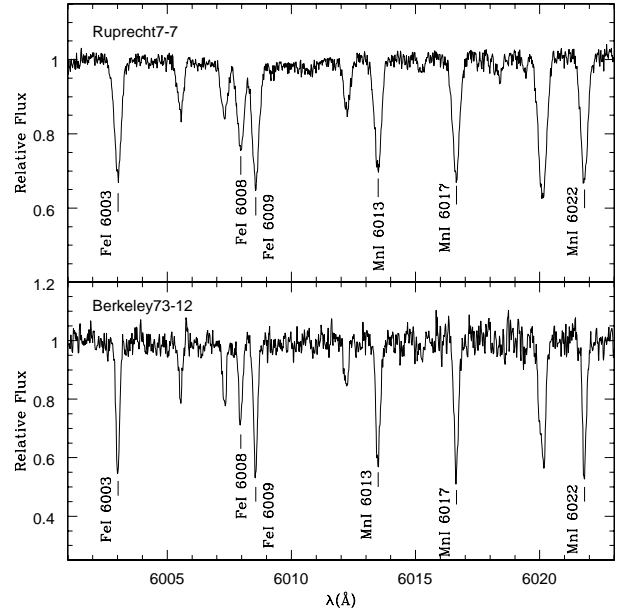}
      \caption{The spectrum of Ruprecht~7-7 (upper panel) and
	Berkeley~73-12 (lower panel) in the wave-length interval
        6000-6025 \AA\ . Several
	spectral lines are identified. }
         \label{Figgrad}
   \end{figure}

\section{Observations and data reduction}
The spectroscopic data come from the ESO-VLT ,
and were collected in service mode between October and December 2005
with the FLAMES on-board VLT\-+UVES
high-resolution, echelle spectrograph.  The sky
was generally clear, and the typical seeing was about 0.8 arcsec during
all of the observing nights.  We used the UVES
multifiber mode, which allows the collection of up to eight spectra simultaneously.
The 580nm set-up ($R=40000$ in the  4750-6800 \AA \ range) was used.
The UVES fibers
were placed on the brighter probable members
of the cluster according to their CMD position.
The data were reduced by ESO personnel using the FLAMES-UVES pipeline\footnote{See 
http://www.eso.org/projects/dfs/dfs-shared/web/vlt/vlt-instrument-pipelines.html
for documentation on the FLAMES-UVES pipeline and software.}
which first corrects the spectra for bias and flatfield. Next
a wavelength calibration based on Th-Ar calibration-lamp spectra
was applied. Finally, each spectrum is flux-calibrated applying
the response-curve of the instrument and the echelle orders combined
to obtain a single one-dimensional spectrum.
The resulting spectra have a dispersion of 0.1 \AA/pixel and a typical
$S/N\sim$ 25--40.
We could observe 21 Red Giant Branch(RGB)/red clump stars in the five program clusters,
for a grand total of 43 spectra. Due to the small size of the
clusters under study,
it was not possible to put all eight fibers on cluster
stars. Typically we could configure UVES to pick up five stars
per cluster.  In the case of Berkeley~75 we could observe just two
stars with sufficiently high $S/N$.
Table~2 provides information on the targets including photometry from
Carraro et al. (2005a,b).
Note that we obtained 1--3 independent spectra (n, column 9 in Table~2) per star, and
the integration time was 45 minutes per exposure.
Examples of extracted spectra are illustrated in Figure 1, where the
spectrum of star $\#7$ in Ruprecht~7 and $\#12$ in Berkeley~73 are shown.

\subsection{Radial velocities}
We used the {\it fxcor} IRAF utility  to measure radial velocities.
This routine cross-correlates the observed spectrum with a template
having known radial velocity.
As a template we used a synthetic spectrum calculated
for a typical solar metallicity giant star ($T_{\rm{eff}}\sim5000\ K$,
$log(g)=3.0$, $v_t=1.3\ km/s$).
Then each measured radial velocity was converted to heliocentric velocity
using the {\it rvcorr} IRAF routine.
The resulting heliocentric velocities for the stars ($RV_{\rm{H}}$)
are reported in Table~2. The error in radial
velocity - derived from repeated measures - is less than 1 km s$^{-1}$.
Finally for the abundance analysis each spectrum was shifted to rest-frame velocity
and continuum-normalized.

\addtocounter{table}{1}

\section{Abundance analysis}

\subsection{Atomic Parameters and Equivalent Widths}

The analysis of chemical abundances was carried out with the 2005
version of the freely available program {\it MOOG} developed by Chris
Sneden\footnote{http://verdi.as.utexas.edu/moog.html}
and originally described in Sneden (1973) and using model
atmospheres by Kurucz (1992). 
{\it MOOG} performs a local thermodynamical
equilibrium (LTE) analysis.
We derived equivalent widths
by fitting Gaussian profiles to spectral lines. Repeated measurements show a
typical error of about 5-10m~\AA\   for the weakest lines
because of the moderate $S/N$ of the spectra.
The line list (see Table~3) was
taken from Gratton et al. (2003). 
The log($gf$)
parameters of these lines were redetermined by a solar-inverse analysis
measuring the equivalent widths from the NOAO  (Kurucz et al. 1984)
solar spectrum, adopting the
standard solar parameters ($T_{\rm{eff}} = 5777$ K, $\log(g)$ = 4.44,
$v_t = 0.8$ km s$^{-1}$, A(Fe) = 7.48).

\subsection{Atmospheric Parameters}
Initial estimates of the atmospheric parameter $T_{\rm{eff}}$ were
obtained from the photometric observations given in Table~2 using the relations from Alonso et al. (1999).
We adopted $E(B-V)$ values from Carraro et al. (2005a,b) to correct colours for the interstellar extinction.
We then adjusted the effective temperature by minimizing the slope of
the abundances obtained from  Fe I lines with respect to the excitation
potential.
Initial guesses for the gravity $\log(g)$ were derived from the canonical
formula:

\begin{equation}
\log\left(\frac{g}{g_{\odot}}\right) =
\log\left(\frac{M}{M_{\odot}}\right) + 4
\log\left(\frac{T_{\rm{eff}}}{T_{\odot}}\right)
- \log\left(\frac{L}{L_{\odot}}\right)
\end{equation}

\noindent
In this equation, the mass $M/M_{\odot}$ was derived from the
comparison between the position of the star in the Hertzsprung-Russell
diagram and the Padova isochrones (Girardi et al. 2000).
The luminosity $L/L_{\odot}$ was derived from the apparent magnitude $V$,
assuming the distance moduli from Carraro et al. (2005a,b)
The bolometric correction
(BC) was derived from the BC-$T_{\rm eff}$ relation from Alonso et al. (1999).
The input $\log(g)$ values were then adjusted to satisfy the
ionization equilibrium of Fe I and Fe II during the abundance
analysis.
Finally, the micro-turbulence velocity is given by the relation (Houdashelt et al.\ 2000):
$v_{\rm t} = 2.22 - 0.322 \log g$.
We then adjusted the micro-turbulence velocity by minimizing the slope
of the abundances obtained from Fe I lines.
The adopted values for all these parameters together with derived [Fe/H] are reported
in Table~4,  whereas the mean results of the detailed abundance analysis per cluster (i.e. with results
averaged over all adopted cluster members --- see \S5) are listed
in Table~5.  We also report there
our measures of the main $\alpha$ elements with their uncertainties and
abundance ratios.

We made a detailed error analysis and found that an increase of 0.1 in
micro-turbulence velocity
implies an increase of 0.01 dex in any of the measured elements (Fe, Mg,
Si, Ca and Ti). On the other hand, a variation of 0.2 in $\log(g)$
produces a decrease of 0.01 dex in any of the elements.
Much more sensitive is the dependence on temperature. An increase
of 100$^{o}$K produces variations as large as +0.11, +0.08, +0.06, +0.10,
and +0.16 dex in Fe, Mg, Si, Ca, and Ti, respectively.\\

\begin{table}
\fontsize{8} {10pt}\selectfont
\tabcolsep 0.1truecm
\caption{Atmospheric parameters. Typical errors
in temperature, logarithm of gravity and
micro-turbulent velocity are 100 $^{o}$K, 0.2 , and 0.1
km/sec.}
\begin{tabular}{cccccccc}
\multicolumn{1}{c}{Cluster} &
\multicolumn{1}{c}{ID} &
\multicolumn{1}{c}{T$_{eff}$} &
\multicolumn{1}{c}{$\log(g)$} &
\multicolumn{1}{c}{$v_t$} &
\multicolumn{1}{c}{log(Fe)} &
\multicolumn{1}{c}{$[Fe/H]$} &
\multicolumn{1}{c}{Membership}\\
\hline
& & $^o$K & & km s$^{-1}$  & & dex & \\
\hline
Berkeley~75 & 9 &4968 &2.57 &1.55   &7.04  &-0.44$\pm$0.03&NM\\
            &22 &5180 &3.37 &1.21   &7.26  &-0.22$\pm$0.03&M\\
Berkeley~25 &10 &5000 &2.90 &1.65   &7.58  &+0.10$\pm$0.05&NM\\
            &12 &4870 &2.75 &1.50   &7.28  &-0.20$\pm$0.04&M\\
            &13 &4860 &2.65 &1.73   &7.31  &-0.17$\pm$0.04&M\\
            &21 &5100 &2.50 &1.80   &7.32  &-0.19$\pm$0.14&M\\
            &23 &5000 &2.50 &1.80   &7.26  &-0.25$\pm$0.12&M\\
Ruprecht~7  & 2 &5160 &2.12 &1.62   &7.14  &-0.34$\pm$0.02&M\\
            & 4 &5105 &2.05 &1.90   &7.24  &-0.24$\pm$0.01&M\\
            & 5 &5230 &2.19 &2.10   &7.21  &-0.27$\pm$0.01&M\\
            & 6 &5230 &2.23 &2.08   &7.28  &-0.20$\pm$0.01&M\\
            & 7 &5150 &2.40 &1.82   &7.23  &-0.25$\pm$0.01&M\\
Ruprecht~4  & 3 &5180 &2.63 &1.56   &7.41  &-0.07$\pm$0.01&M\\
            & 4 &5150 &2.52 &1.66   &7.45  &-0.04$\pm$0.02&M\\
            & 8 &5190 &2.64 &1.40   &7.32  &-0.16$\pm$0.01&M\\
            &18 &5040 &3.17 &1.20   &7.13  &-0.35$\pm$0.01&NM\\
            &29 &4920 &2.78 &1.37   &7.14  &-0.34$\pm$0.02&NM\\
Berkeley~73 &12 &5030 &2.78 &1.40   &7.09  &-0.39$\pm$0.02&NM\\
            &13 &5730 &4.15 &0.99   &7.65  &+0.17$\pm$0.01&NM\\
            &15 &5070 &3.12 &1.04   &7.10  &-0.38$\pm$0.01&NM\\
            &16 &4890 &2.71 &1.45   &7.30  &-0.18$\pm$0.02&M\\
            &18 &4940 &2.88 &1.32   &7.23  &-0.27$\pm$0.02&M\\
            &19 &5870 &4.23 &1.40   &7.45  &-0.03$\pm$0.03&NM\\
\hline
\end{tabular}
\end{table}

%----------------------------------------------------------- S_vib

   \begin{figure*}
   \centering
   \includegraphics{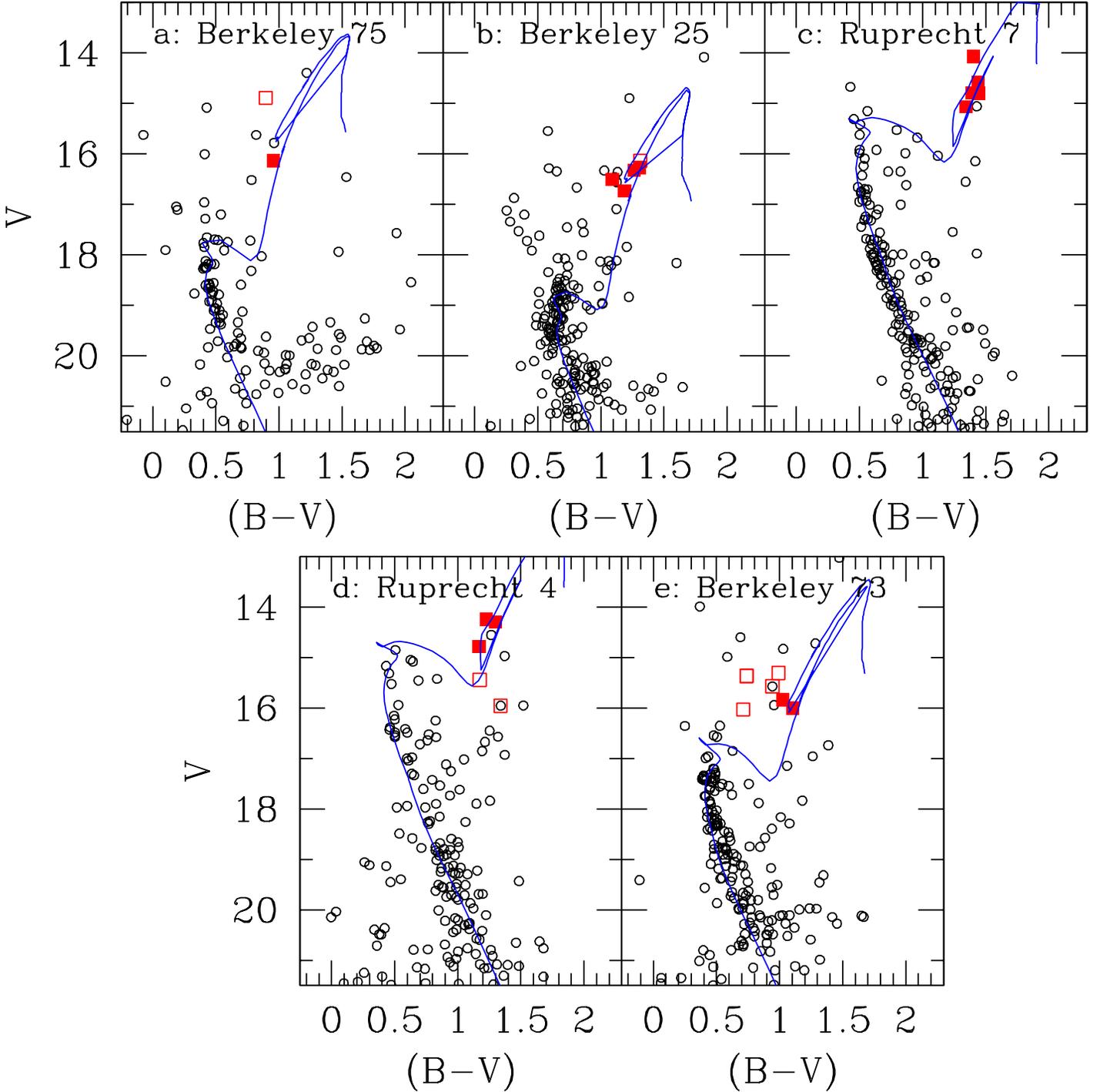}
      \caption{Revision of clusters' fundamental parameters. In each
              panel
              we show the isochrone
              solutions and indicate radial velocity members (filled
              squares) and non-members (empty squares) for (a) Berkeley~75, (b) Berkeley~25, (c) Ruprecht~7,
              (d) Ruprecht~4, and (e) Berkeley~73, respectively.}
         \label{Figcmd}
   \end{figure*}
%
%______________________________________________________________

\begin{table*}
\caption{Mean abundance analysis from cluster members. In
  parenthesis  below the element the number of lines used is indicated}
\fontsize{8} {10pt}\selectfont
\begin{tabular}{cccccccccccc}
\hline
\multicolumn{1}{c} {$Cluster$} &
\multicolumn{1}{c} {$[Fe/H]$} &
\multicolumn{1}{c} {$[Mg/H]$}&
\multicolumn{1}{c} {$[Si/H]$}  &
\multicolumn{1}{c} {$[Ca/H]$} &
\multicolumn{1}{c} {$[Ti/H]$} &
\multicolumn{1}{c} {$[Mg/Fe]$} &
\multicolumn{1}{c} {$[Si/Fe]$} &
\multicolumn{1}{c} {$[Ca/Fe]$} &
\multicolumn{1}{c} {$[Ti/Fe]$} &
\multicolumn{1}{c} {$[\alpha/Fe]$}\\
\hline
 & (150) & (3) & (13) & (16) & (33) & &&&&\\
\hline
Berkeley~75& -0.22$\pm$0.20 & +0.16$\pm$0.06 & -0.42$\pm$0.14 & -0.24$\pm$0.07& -0.06$\pm$0.04& +0.38$\pm$0.09 & -0.20$\pm$0.17&-0.02$\pm$0.10 &+0.16$\pm$0.07&+0.08\\
Berkeley~25& -0.20$\pm$0.05 &  0.00$\pm$0.29 & -0.12$\pm$0.17 & -0.30$\pm$0.10& -0.07$\pm$0.13& +0.11$\pm$0.31 & -0.01$\pm$0.19&-0.19$\pm$0.12 &+0.04$\pm$0.15&-0.01\\
Ruprecht~7 & -0.26$\pm$0.05 & -0.35$\pm$0.09 & -0.20$\pm$0.05 & -0.22$\pm$0.11& -0.37$\pm$0.09& -0.10$\pm$0.10 & +0.05$\pm$0.06&+0.03$\pm$0.12 &-0.12$\pm$0.10&-0.04\\
Ruprecht~4 & -0.09$\pm$0.05 & -0.09$\pm$0.04 & -0.05$\pm$0.04 & -0.05$\pm$0.05& -0.06$\pm$0.05& +0.01$\pm$0.05 & +0.05$\pm$0.05&+0.05$\pm$0.06 &+0.04$\pm$0.06&+0.04\\
Berkeley~73& -0.22$\pm$0.10 & -0.12$\pm$0.10 & -0.05$\pm$0.04 & -0.30$\pm$0.07& -0.17$\pm$0.07& +0.10$\pm$0.15 & +0.17$\pm$0.09&-0.08$\pm$0.19 &+0.05$\pm$0.12&+0.06\\
\hline
\end{tabular}
\end{table*}

\section{Revision of cluster fundamental parameters}
We make use here of the derived radial velocity, metallicity,
and available photometry to revise the fundamental parameters of the clusters under
investigation.

First, we discuss
the membership of all the program stars on the basis of their radial
velocity (Table~2), position in the CMD,
location in the cluster field, atmospheric parameters, and chemistry.  From
the stars that have been selected as members,
we compute the cluster mean metal abundance (see Table~5).
We use the following criterion to add error bars to the cluster mean
iron abundance [Fe/H]. 
We adopt as metallicity error 0.2 dex when cluster metallicity is
based on just 1 member, 0.1 dex in the case we find two members, and
0.05 dex when we have 3 or more members, but only in the case the
measurements standard deviation $\sigma$
is smaller than this value, otherwise we directly
adopt
the measurements $\sigma$.
 
We then generate isochrones for the
derived spectroscopic metallicity of each cluster, transforming the mean [Fe/H] into $Z$,
using Padova models and following Carraro et al. (1999).
Owing to the marginal $\alpha-$element enhancement, we derived
$Z$ considering only [Fe/H].
The corresponding best-fit isochrone is then superimposed on each CMD
(Fig.~2).
In this way updated estimates
of the basic parameters (age, distance and reddening) are finally
derived (see Table~6). Distances are computed adopting $A_V = 3.1 \cdot
E(B-V)$ (Moitinho 2001).
Finally, by assuming 8.5 kpc as the Sun's distance from the Galactic
center,
we provide in Table~7 the Cartesian Galactic coordinates $X_{GC}$, $Y_{GC}$ and $Z_{GC}$,
and the distance ($R_{GC}$) of the clusters from the Center of the
Galaxy. The Cartesian coordinates are defined as $Z$ pointing toward the
North Galactic Pole, $Y$ toward the direction of the Galactic rotation, and,
finally, $X$ pointing toward the anti-center.

The fundamental parameters are thus summarized in Tables~5, 6 and 7.

\subsection{Berkeley~75}
For this cluster we make use of the $BVI$ photometry
from Carraro et al. (2005a). We have only two stars with spectroscopy and
their velocities are clearly incompatible (see Table~2), as are
their metallicities (Table~4).
In Fig.~2 (panel a) we have indicated these two stars with squares.
The decision on the possible membership of these stars is obviously
quite difficult. In the figure we show as a member (filled square)
the fainter ($\#22$) of the two stars, and suggest that star $\#9$ (open square)
is actually an interloper.
First,  star $\#22$ falls closer to the adjusted isochrone in the CMD, while star
$\#9$ seems too bright and blue.
In addition, in the figure (where only the stars within the cluster
radius are shown),
while star $\#22$ lies within the cluster radius,
star $\#9$ is close to the edge of the field of view,
well outside the cluster radius; this star is shown in Figure 2a
only because it was observed spectroscopically.
Finally, from Table~2
one can see how star $\#9$ is actually  cooler than star $\#22$.
If one derives their reddening from the intrinsic colors
(Worthey et al. 2006), it turns out that  star  $\#22$
has a reddening compatible with the bulk of the MS stars (0.04),
while star $\#9$ basically does not have reddening, being
therefore a nearby giant star. The probable member ($\#22$)
has a metallicity [Fe/H]  $= -0.22\pm0.20$, and this translates into
$Z = 0.011$ (Carraro et al. 1999). A 4 Gyr isochrone for this
metallicity has been super-imposed in Figure 2a shifted by
$E(B-V) = 0.04$ and $(m-M)_V$ = 14.9. Note that our inferred distance
of this cluster turns out to be lower than previous estimates (Table~1, Carraro et
al. 2005a) due to the much higher metallicity we derived from spectroscopy.
However, we consider our derived values
for this cluster to be very preliminary
since they are in part  based on only one star
with ambiguities in determining its membership.

\subsection{Berkeley~25}
The photometry for this cluster is also taken from Carraro et al. (2005a).
We have five stars with very similar radial velocities, and therefore
all of the measured stars are probable members.
However, only for three of them ($\#$10, 12, and 13) could we
perform detailed abundance analysis, while for the other two only an
estimate of the iron abundance was possible, due to the lower $S/N$.
These [Fe/H] values are comparable within the error among all 
members, except for star $\#10$, which we are here considering as a
probable non-member.
These stars are indicated with
filled squares in Figure 2b. The mean of the four
metallicity measures (Table~4) yields [Fe/H]= -0.20$\pm$0.05,
which implies $Z = 0.012$.
The best-fit is for a 5 Gyr isochrone, and it
is shown in the same panel.
It implies $E(B-V) = 0.11$ and $(m-M)_V$=15.6.

\subsection{Ruprecht~7}
The photometry of this cluster comes from Carraro et al. (2005b).  We obtained
spectra for five stars (Table~2) and all have compatible radial
velocities and metallicities within the errors. The photometry of these stars is indicated
in Figure 2c.  The mean of the five
metallicity measures yields [Fe/H]= -0.26$\pm$0.05, which corresponds
to $Z = 0.010$.
The 800 Myr isochrone that provides the best fit
for this metallicity has been
drawn for $E(B-V) = 0.36$ and $(m-M)_V$=15.0.
Owing to the lower metallicity we measure,
these findings somewhat deviate  from  the purely
photometric estimates by Carraro et al. (2005b).

\subsection{Ruprecht~4}
Of the five stars with Carraro et
al. (2005b) photometry for which we obtained spectra, we propose that three ($\#$ 3, 4 and 8)  are
cluster members (filled squares in Fig.~2d), while the other two (the two faintest)
are interlopers. This choice is also motivated by the better position of the
candidate member stars in the CMD (see Fig.~2d).  Four of the
five stars lie near the cluster center (1.5 arcmin,
Carraro et al. 2005b), but star $\#$29
is located well outside, suggesting it is a probable non-member.
This further corroborates our member selection.
From the three adopted members, we derive a mean
[Fe/H]= -0.09$\pm$0.05, hence $Z = 0.015$.
The 800 Myr isochrone for this metal content has been
superimposed using $E(B-V) = 0.30$ and $(m-M)_V$=14.30.

\subsection{Berkeley~73}
The photometry of this cluster comes from Carraro et al. (2005a).
In this case assigning spectroscopically-based memberships is again quite a difficult task.
Among the six stars for which we obtained UVES spectra,
we do not consider the two stars with dwarf-like gravities, $\#$13 and 19, as members,
while we consider the stars $\#$16 and 18 as members, because they are
both giants with comparable velocities and metallicities.
Star  $\#$15 is a probable radial velocity non-member, as is star
$\#$12.
We therefore are left with two stars we consider to be members, and their
mean [Fe/H] is  -0.22$\pm$0.05.  This translates to $Z = 0.011$,
as in the case of Berkeley~75.
The 1.5 Gyr best fit
isochrone shown in Figure 2e has been shifted by assuming $E(B-V) =
0.11$ and $(m-M)_V$ = 15.30, a result consistent with the purely photometric
analysis by Carraro et al. (2005a).

\begin{table*}
\caption{Revision of the fundamental parameters of the clusters under study.}
\fontsize{8} {10pt}\selectfont
\begin{tabular}{cccccccc}
\hline
\multicolumn{1}{c} {$Name$} &
\multicolumn{1}{c} {$[Fe/H]$} &
\multicolumn{1}{c} {Metallicity}&
\multicolumn{1}{c} {$E(B-V)$}  &
\multicolumn{1}{c} {$(m-M)$} &
\multicolumn{1}{c} {$d_{\odot}$} &
\multicolumn{1}{c} {$Age$} \\
\hline
           & dex            &       &               &      &  kpc & Gyr\\
\hline
Berkeley~75& -0.22$\pm$0.20 & 0.011 & 0.04$\pm$0.03 & 14.90$\pm$0.20 & 9.10 & 4.0$\pm$0.4\\
Berkeley~25& -0.20$\pm$0.05 & 0.012 & 0.11$\pm$0.05 & 15.60$\pm$0.30 &11.40 & 5.0$\pm$0.5\\
Ruprecht~7 & -0.26$\pm$0.05 & 0.010 & 0.36$\pm$0.05 & 15.00$\pm$0.20 & 6.00 & 0.8$\pm$0.2\\
Ruprecht~4 & -0.09$\pm$0.05 & 0.015 & 0.30$\pm$0.05 & 14.30$\pm$0.20 & 4.70 & 0.8$\pm$0.2\\
Berkeley~73& -0.22$\pm$0.10 & 0.011 & 0.11$\pm$0.05 & 15.30$\pm$0.20 & 9.80 & 1.5$\pm$0.3\\
\hline
\end{tabular}
\end{table*}

\begin{table*}
\caption{Revision of the fundamental parameters of the clusters under study.}
\fontsize{8} {10pt}\selectfont
\begin{tabular}{ccccccccccc}
\hline
\multicolumn{1}{c} {$Name$} &
\multicolumn{1}{c} {$l$} &
\multicolumn{1}{c} {$b$} &
\multicolumn{1}{c} {$[Fe/H]$} &
\multicolumn{1}{c} {$X$} &
\multicolumn{1}{c} {$Y$} &
\multicolumn{1}{c} {$Z$} &
\multicolumn{1}{c} {$R_{GC}$} &
\multicolumn{1}{c} {$RV$} &
\multicolumn{1}{c} {$V_{GSR}$} &
\multicolumn{1}{c} {$V_{ROT}$}\\
\hline
 & [deg] & [deg]  & dex  &  kpc & kpc & kpc & kpc & km/sec & km/sec & km/sec \\
\hline
Berkeley~75& 234.30 & -11.12 &-0.22$\pm$0.20 & 13.7 & -7.2 & -1.6 & 15.5&   +94.6$\pm$0.35& -95.7&220.5\\
Berkeley~25& 226.60 &  -9.69 &-0.20$\pm$0.05 & 16.2 & -8.1 & -1.9 & 18.2&  +134.3$\pm$0.20& -39.1&117.4\\
Ruprecht~7 & 225.44 &  -4.58 &-0.26$\pm$0.05 & 12.7 & -4.6 & -0.5 & 13.5&   +76.6$\pm$0.50& -95.0&210.9\\
Ruprecht~4 & 222.04 &  -5.31 &-0.09$\pm$0.05 & 12.0 & -3.1 & -0.4 & 12.4&   +47.5$\pm$1.00&-114.5&250.2\\
Berkeley~73& 215.28 &  -9.42 &-0.22$\pm$0.10 & 16.4 & -5.6 & -1.6 & 17.4&   +95.7$\pm$0.20& -44.9&155.0\\
\hline
\end{tabular}
\end{table*}

\section{Extension of the cluster sample}
Out of 5 total, 3 of our sample (the Berkeley clusters) are at heliocentric distances
larger than 7 kpc, and therefore have to  be carefully
considered as possible previously neglected MRi candidates (Crane et
al. 2003, Frinchaboy et al. 2004). 
Therefore they deserve further attention to
better describe this tidal feature.
At the same time, we want to investigate the chemical and kinematic properties of the
outer Galactic disk using old open clusters.\\
\noindent
To these aims,  we enlarge the sample presented in the previous Sections
by adding eight additional distant anti-center clusters for
which high resolution abundances are available.
These clusters are: Berkeley~29
and Saurer~1 (Carraro et al. 2004), Berkeley~22 (Villanova et
al. 2005), Tombaugh~2 (Frinchaboy et al. 2007),  Berkeley~21 (Hill \&
Pasquini 1999; Yong et al. 2005), and Berkeley 20,
Berkeley 31, and NGC~2141 from Yong et al. (2005).
These additional clusters are all located in the Third Galactic Quadrant, having
$187^{\circ} \leq l \leq 240^{\circ}$ and $-18^{\circ} \leq b \leq +8^{\circ}$.  Their properties
are summarized in Table~8.
These thirteen clusters all lie more
than 12 kpc from the Galactic center and represent the largest
sample of anti-center distant clusters with reliable
chemical abundance measurements analyzed together so far.
By the way,  3 of them (Berkeley~29, Saurer~1 and Tombaugh~2) have been
previously suggested to be MRi candidates.\\

\noindent
In order to put all the cluster [Fe/H] measurements taken from
  the literature on the
  same metallicity scale as the present study clusters, we 
  used the values for Berkeley 29, which is in common between Yong et
  al. (2005) and Carraro et al. (2004). This allows us to scale Berkeley~20, Berkeley~31 and NGC~2141, 
  and Berkeley~21. For this latter we still consider the Yong et al.
  value, since it is basically identical to Hill \& Pasquini (1999)
  estimate. For Berkeley~29, Yong et al. report an [Fe/H] value 0.1 dex
  lower than Carraro et al. (2004). Therefore we increase all Yong et
  al.  [Fe/H] values by 0.1 dex.

    \begin{figure}
   \centering
   \includegraphics[width=\columnwidth]{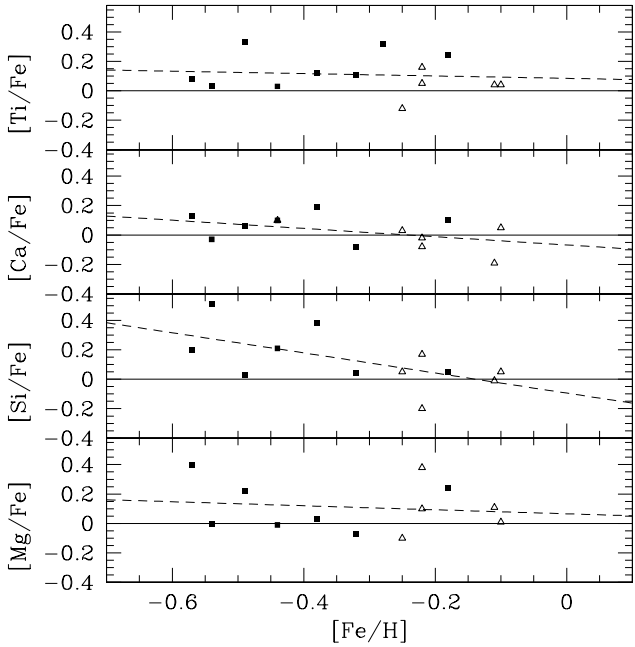}
      \caption{The trend of $\alpha$ element abundances with metallicity. Open triangles
	indicate the five program clusters, while filled squares are additional
	anti-center clusters culled from the literature.The dashed line is
      a fit through the data.}
         \label{Figgrad}
   \end{figure}

\section{Abundance ratios}

Carraro et al. (2004) and Yong et al. (2005) have emphasized that
the few outer Galactic disk clusters insofar studied show a sizeable enhancement in  $\alpha$
elements over the solar ratio.
Having a larger sample, we investigate here what the $\alpha$
element ratios are suggesting to us about the origin of these clusters,
say whether they are genuine disk clusters, or whether they are
the signature of an extra-galactic population.
Indeed one may  speculate that an $\alpha$ element over-abundance
indicates that the  material in the outer
disk formed rapidly, possibly  due to a relatively recent merging
event (2-5 Gyrs ago).
This seems to be confirmed by the ages of the clusters under analysis.
None of the clusters in our sample is very old (ages less than 5 Gyr),
which  may suggest that very old open clusters (like NGG 6791, Berkeley 17,
Collinder 261 and  NGC 188 to name a few examples) do not
populate the anti-center. We emphasize however that this might
be the result of a selection effect, which further studies have to clarify.
Nonetheless, if this age distribution is real, one may speculate that these
clusters
could have formed during a merger, or deposited by a merger.
In fact, according to our current understanding of disk formation,
we expect that the outer parts have been forming later (inside
out scenario) and that the more recently accreted material, if any, is around
the outer edge of the disk.\\

\noindent
We could measure several $\alpha$ elements for our five clusters,
as listed in Table~5.  Figure~3 shows the [Fe/H] trends of
four abundance ratios: [Mg/Fe], [Si/Fe], [Ca/Fe], and [Ti/Fe] of our
5 program clusters, plus the 8 additional clusters collected from the
literature.
The dashed lines are linear fits through the data points with slopes of
-0.14$\pm$0.11, -0.58$\pm$0.19, -0.28$\pm$0.12, and -0.08$\pm$0.10 for
[Mg/Fe], [Si/Fe], [Ca/Fe], and [Ti/Fe], respectively.
There is a trend of having  decreasing $\alpha$ elements ratios
at increasing metallicity ([Fe/H]) for all the measured
elements, except for Ti.

Inspecting this figure we notice  that these clusters define
a trend of $\alpha$ element abundances versus [Fe/H] consistent with Galactic
open clusters (Friel et al. 2003, Fig.~5) and field F \& G stars (Reddy et
al. 2003, Fig.8) near the Sun in the same metallicity range
of our sample.
The generally enhanced abundances are more reminiscent of the
thick than thin disk behavior (Reddy et al. 2003).

At the metallicity there are no dwarf galaxies having the same
abundance ratios. The Sagittarius Dwarf Spheroidal does reach
these high metallicities ($-0.5\leq [Fe/H] \leq 0.0$), however
its stars tend to be significantly under-abundant in $\alpha$ elements.
The unpublished work by Smecker-Hane \& McWilliam ({\tt astro-ph/
0205411}) present a few determinations compatible with our
clusters. However, more recent studies (Sbordone et al. 2007, and
references therein) based on a larger sample show that in this
metallicity range $\alpha$ elements are significantly sub-solar.

In conclusion, the $\alpha$ element ratios of outer disk
open clusters are  more compatible with the solar vicinity 
thin/thick disk than
with dwarf Galaxies, suggesting that these clusters are genuine
members of the Galactic disk.

  \begin{figure}
   \centering
   \includegraphics[width=\columnwidth]{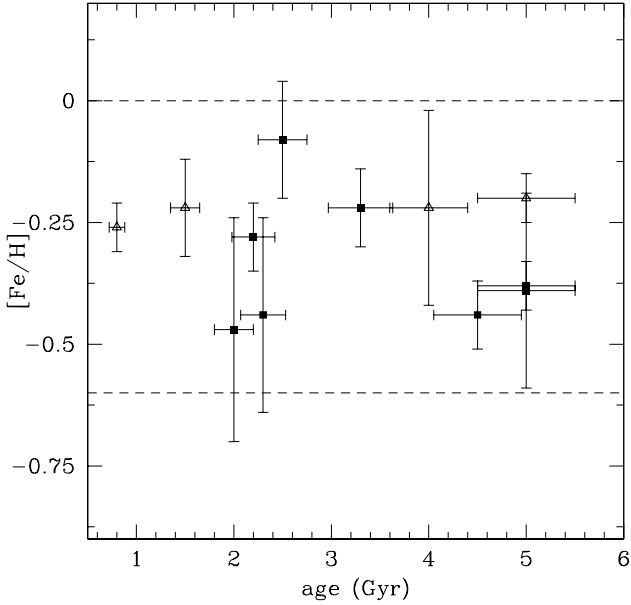}
      \caption{Age metallicity distribution, with symbols as in
      Fig.~3. The dashed lines indicate the metallicity range of old
      open clusters in the Galactic disk, from Friel et al. 2002.}
         \label{Figgrad}
   \end{figure}

\section{Age-Metallicity Relationship}
The age-metallicity relationship (AMR) is another important diagnostic of
galaxy formation scenarios.  If our clusters were born in a distinct galaxy,
they would likely show an AMR distinct from the clusters intrinsic to our Galaxy.

For example, Frinchaboy et al. (2004)
suggest the open and globular clusters putatively associated with MRi show an
AMR 
similar to that of the Sagittarius dwarf galaxy (Layden \& Sarajedini 2000).
Forbes et al. (2004) built on this concept and used the
AMR as a tool to decide the membership or not of potential
CMa/MRi clusters, as did Frinchaboy et al. (2006).
However, both these studies are based on highly inhomogeneous samples,
combine together open and globular clusters,
and make use of photometry- or low-resolution-sprectroscopy-based metallicities.

It is quite well accepted that the bulk of the old open clusters
in the disk do not define a clear-cut AMR
(Friel 1995, Carraro et al. 1998, Friel et al. 2002), but rather at any
age these clusters show a large scatter in metallicity.
Typically, at any age between 1 and 8 Gyr, metallicity ranges from
[Fe/H] $\sim$ -0.5 to [Fe/H] $\sim$ +0.05 (Friel et al. 2002, Fig.4).

This seems to be the case also for outer disk clusters (see Fig.~4):
At all ages from 0.8 to 5 Gyr there is a sizable spread in [Fe/H]
extending up to 0.4 dex, in good agreement with the trend of
old open clusters in the solar neighborhood. The two lines
in Fig.4 show the full metallicity range that old open clusters in the
disk exhibit at any age (Friel et al. 2002).

There might be a possible hint of an AMR if we remove the two most
metal-poor clusters
(Berkeley~21 and NGC 2141, Yong et al. 2005) located at about 2 Gyrs.
However, despite the large errors, the metallicity of these two clusters
are consistent with all the determinations obtained so far (Hill
\& Pasquini 1999), and there are no reason to believe they are wrong.

Therefore, the analysis of the AMR lends further support to the idea
that the outer disk old open clusters do not differ from typical disk
clusters, since they do show the same lack of any AMR as typical disk open
clusters and argue against their origin in a putative dwarf galaxy.

\section{Metallicity distribution}
The sample of clusters we are analyzing has a mean metallicity
[Fe/H] =  -0.34$\pm$0.15. It is useful to compare this
value with the available measures for the MRi.
Yanny et al. (2003) derived [Fe/H]=-1.6$\pm$0.3 from Ca II (K)
line strenghts of MS blue F/G star spectra.
This metallicity has been recently confirmed by Wilhelm et al. (2005),
who used synthetic spectra and colors to derive 
[Fe/H]=-1.37$\pm$0.04 from a group
of A/F stars.
From the spectra of a sample of M giants Crane et al. (2003) obtained
quite a different result. The Ca II $\lambda\lambda$8498, 8542 and Mg I
$\lambda$8807 lines synthesis yielded [Fe/H] =-0.4$\pm$0.3.
If the F/G MS stars and the M giants belong to the same
population, they seem to define an AMR, provided they are
sampling different age sub-populations, corresponding to different
rounds of star formation.
If, on the contrary, F/G and M stars
do belong to the same population and possess roughly the same age,
clearly something is wrong with these estimates of the metallicity.

Interestingly, our clusters do have the same metallicity as the 
Crane et al. (2003) M giants.
One may be tempted to associate old open clusters with M giants, while
associating putative members globular clusters (Frinchaboy et
al. 2004) with F/G stars.
While in principle this might be correct - the Sgr DSph represents
a clear counterpart -
such a conclusion still remains premature,  first because $\alpha$
elements and AMR are suggesting a different scenario,
second because the MRi global properties are still very poorly known.

Besides, one is left with the question as to why outer disk star
clusters should have roughly the same metallicity distribution as solar
vicinity star clusters (Friel et al. 2002).

\begin{table*}
\caption{Additional clusters gathered from the literature. References are: (a) Carraro et al. (2004); (b) Yong et al. (2005);
(c) Villanova et al. (2005); (d) Frinchaboy et al. (2007).}
\fontsize{8} {10pt}\selectfont
\begin{tabular}{ccccccccccccccc}
\hline
\multicolumn{1}{c} {$Name$} &
\multicolumn{1}{c} {$l$} &
\multicolumn{1}{c} {$b$} &
\multicolumn{1}{c} {$RV_H$} &
\multicolumn{1}{c} {$d_{\odot}$} &
\multicolumn{1}{c} {$R_{GC}$} &
\multicolumn{1}{c} {$V_{ROT}$} &
\multicolumn{1}{c} {$V_{GSR}$}&
\multicolumn{1}{c} {$[Fe/H]$} &
\multicolumn{1}{c} {$[Mg/Fe]$} &
\multicolumn{1}{c} {$[Si/Fe]$} &
\multicolumn{1}{c} {$[Ca/Fe]$} &
\multicolumn{1}{c} {$[Ti/Fe]$} &
\multicolumn{1}{c} {$age$} &
\multicolumn{1}{c} {$Ref$} \\
\hline
 & deg & deg & km/sec & kpc & kpc & km/sec & km/sec &dex & dex & dex & dex & dex & Gyr & \\
\hline
Berkeley~29 & 197.98 &  +8.05 & +24.6 & 13.2 & 21.6 &340.0 & -51.1 &-0.44 & -0.01 & +0.21 & +0.10 & +0.03& 4.5&a)\\
Saurer~1    & 214.68 &  +7.38 &+104.6 & 13.2 & 19.2 &141.6 & -32.7 &-0.38 & +0.05 & +0.38 & +0.19 & +0.12& 5.0&a)\\
Berkeley~31 & 206.25 &  +5.12 & +55.7 &  8.3 & 16.3 &210.9 & -48.6 &-0.57 & +0.40 & +0.20 & +0.13 & +0.08& 2.0&b)\\
Berkeley~22 & 199.88 &  -8.08 & +95.3 &  6.0 & 14.3 & 43.3 &  +7.8 &-0.32 & -0.07 & +0.04 & -0.08 & +0.11& 3.3&c)\\
Berkeley~20 & 203.48 & -17.37 & +78.9 &  8.4 & 16.0 & 99.7 & -19.8 &-0.49 & +0.22 & +0.03 & +0.06 & +0.33& 5.0&b)\\
Tombaugh~2  & 232.00 &  -6.88 &+121.0 &  8.3 & 15.1 &149.6 & -75.8
&-0.28 &    &  &  & +0.32& 2.2&d)\\
Berkeley~21 & 186.84 &  -2.51 & +12.6 &  5.0 & 13.5 &323.0 & -24.2 &-0.54 & +0.00 & +0.51 & -0.03 & +0.03& 2.3&b)\\
NGC~2141    & 198.08 &  -5.78 & +24.1 &  4.3 & 12.0 &177.8 & -37.7 &-0.18 & +0.24 & +0.05 & +0.10 & +0.24& 2.5&b)\\
\hline
\end{tabular}
\end{table*}

   \begin{figure}
   \centering
   \includegraphics[width=\columnwidth]{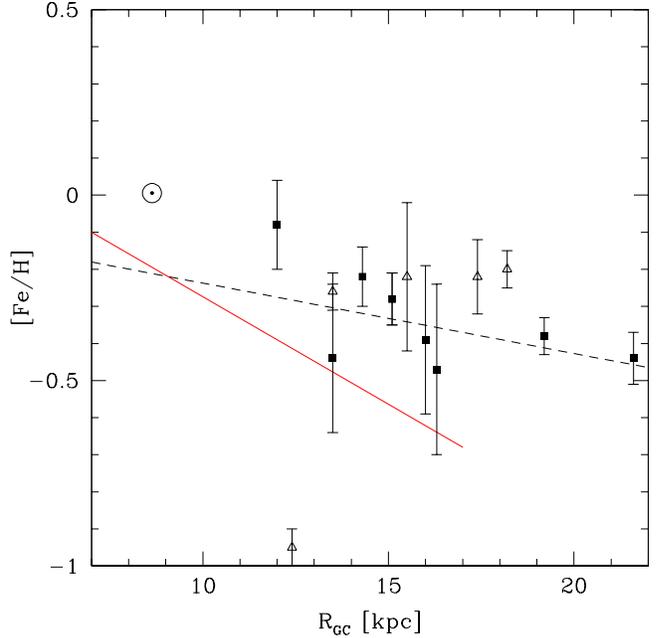}
      \caption{The radial abundance gradient from old open clusters in the distant
anticenter direction. Symbols are as in Fig.~3. The solid line is the Friel
      et al. (2002) mean abundance gradient, while the dashed line is
      the gradient when outer disk clusters are added.
      }
         \label{Figgrad}
   \end{figure}

   \begin{figure}
   \centering
   \includegraphics[width=\columnwidth]{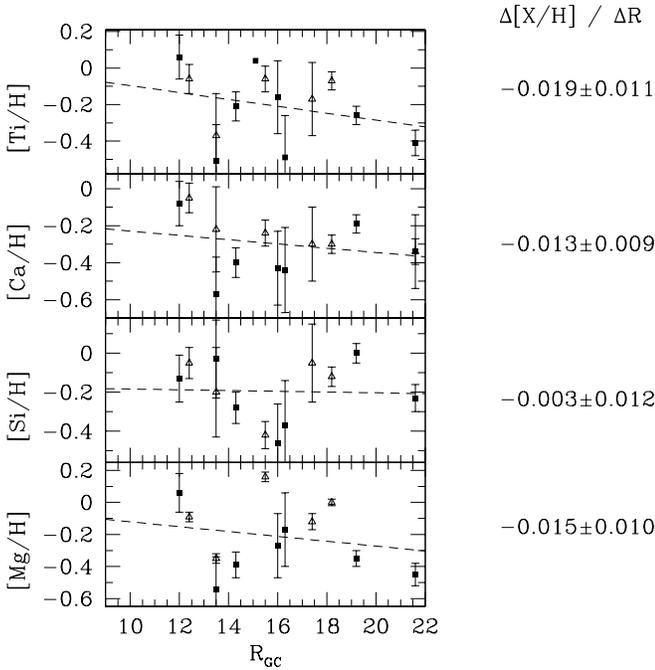}
      \caption{The radial abundance gradient of Mg, Si, Ca and Ti from old open clusters in the distant
       Galactic anti-center. The symbols are the same as those used in Fig.~3. On the right the gradient
       slopes in the range 12 to 22 kpc are indicated.}
         \label{Figgrad}
   \end{figure}

\section{The radial abundance gradient in the outer Galactic disk}
The most recent determination of the slope of the radial abundance
gradient using old open clusters in the Galactic disk is presented in Friel et al. (2002).  From
a sample of about 40 clusters located between 8 and 14 kpc from the
Galactic center a mean slope of -0.06
dex kpc$^{-1}$ is found.
This slope mostly represents the abundance gradient in the solar
vicinity, since the sample contains just one cluster beyond 14 kpc
from the center of the Milky Way and only five clusters outside of 12
kpc.

Carraro et al. (2004) and Yong et al. (2005) provide evidence
that the abundances in the outer disk ($R_{GC}$ larger than 12-13 kpc) do not follow the
expectations of the extrapolation of the
gradient as determined by clusters in the solar
neighborhood, but instead the metallicity gradient seems to flatten at $R_{GC}
\sim 12-14$ kpc.

With the supplement of the eight additional clusters from the literature we
now have a unique sample of clusters
to reassess
the radial abundance gradient in the outskirts of the Galactic disk.
Our sample spans almost 10 kpc in Galactocentric
distance (see Table~5 and 8), from $R_{GC}$ = 12.0 to 21.6 kpc.
Moreover, they are all located essentially toward the same Galactic
region ($186^o \leq l \leq 234^o$).
Figure~5 plots the metallicity trend of all of the aforementioned clusters
together with their metallicity uncertainty.

In this figure, the solid line is the radial abundance gradient
from Friel et al. (2002) sample.
Clearly, distant anticenter old clusters deviate from this trend,
and
show how the mean gradient beyond 12 kpc flattens  out and essentially maintains
a constant value of $[Fe/H] \sim -0.35$.

If we
consider the Friel et al. (2002) clusters and the
ones discussed here together, a global gradient of -0.018$\pm$0.021 dex
kpc$^{-1}$ results (dashed line) --
much flatter than that found by Friel et al. (2002) and
compatible with the Galactic disk not having any gradient when using
old open clusters over its full extent.

We note that shallow metallicity gradients across
the full span of spiral galaxy disks are commonly found
(see Moustakas \& Kennicut 2006, and references therein).\\
\noindent
The global value of the gradient we find is also consistent with that
predicted by Cescutti et al. (2006, see below)
chemical evolution models, which yield a gradient of $\sim $-0.05
dex kpc$^{-1}$ all the way from 4 to 22 kpc from the Galactic center.

In Figure~6 we show the radial abundance gradient of four
$\alpha-$elements --- [Mg/H], [Si/H], [Ca/H], and [Ti/H] --- and
derive their gradient slope (indicated on the right side of each
panel).

Recently, Cescutti et al. (2006) calculated the slopes of the
Galactic abundance gradient of several elements as derived from chemical
evolution models of the Milky Way. This model is based on the
assumption that the MW evolved basically in isolation, but with different
star formation thresholds imposed for
the halo and disk.
The disk is modeled as an inside-out growing structure.
Modern prescriptions for element production rates are employed.

In the same Galactocentric distance baseline (12 to 22 kpc)
the Cescutti et al. models yield abundance gradient slopes of about -0.020 dex kpc$^{-1}$
for the four above elements, while reproducing at the same time the observed gradient
in the solar vicinity.

Therefore, model predictions are consistent with our new results within
the errors, except for [Si/H], for which we found a shallower slope
than the models, consistent with no gradient.  We find that the mean trend of the models
match the trend of the observations for the outer disk.

\section{Spatial distribution and kinematics}
The five program clusters are all located below the nominal Galactic plane and
reach as much as 2 kpc below the plane (see Table~ 7).
In Figure 7 we plot the R$_{GC}$-Z$_{GC}$ distribution of all thirteen clusters
beyond 12 kpc from the Galactic Center collected in Section 5.
In the third Galactic quadrant, stars and clusters
preferentially lie below the formal $b = 0$ plane of the Galaxy,
following the pattern of the Galactic warp and flare (Moitinho et al. 2006). These are shown
for different values of $b$, from 0$^{o}$ (solid line) to $\pm3^{o}$ (dashed lines)
from Momany et al. 2006, which the reader is referred to for any
additional details.
Berkeley~29, Berkeley~31 and Saurer~1  are at odds with the bulk of the sample and
are located well above the Galactic
plane. 
Still, they  do not deviate significantly from the predictions of a warped/flared disk.\\

   \begin{figure}
   \centering
   \includegraphics[width=\columnwidth]{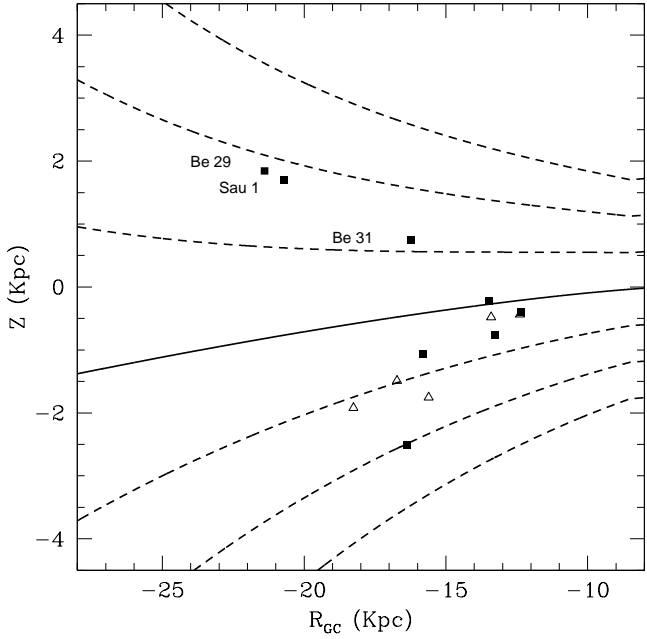}
      \caption{Distribution of star clusters according to their
      height above the Galactic plane and distance from the Galactic
      center. The various lines indicate the warping/flaring of the
      disk for $b=0^{o}$ (solid line) and $\pm$1,2 and 3 scale heights (dashed
      lines). Symbols are as in Fig.~3}
         \label{Figgrad}
   \end{figure}

   \begin{figure}
   \centering
   \includegraphics[width=\columnwidth]{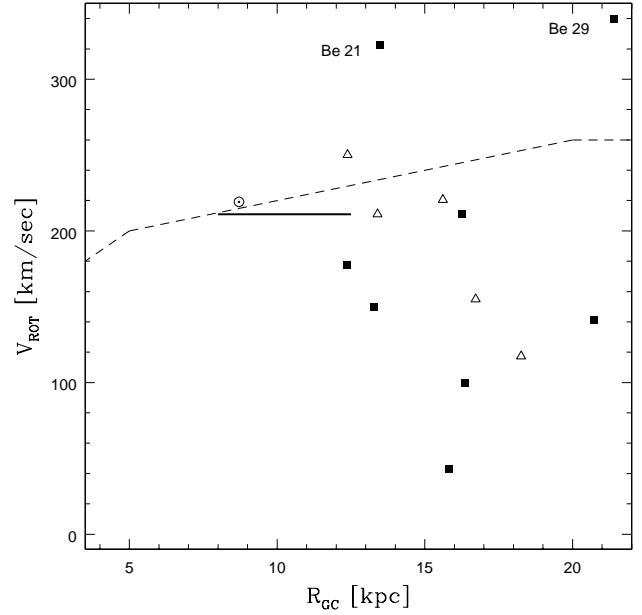}
      \caption{Rotational velocity of old open clusters in the third
      Galactic quadrant beyond 12 kpc from the Galactic center.
      The dashed line is the stellar disk rotation
      curve from Olling and Merrifield (2000), while the thick solid
      line is the typical location of old open clusters in the solar
      vicinity. Symbols are as in Fig.~3}
         \label{Figgrad}
   \end{figure}

\noindent
Scott et al. (1995) have studied the kinematics of old open clusters
in the solar vicinity. From a sample of 35 clusters at mean
Galactocentric distance of 10.3 kpc, they found a mean rotational
velocity of 211$\pm$7 km/sec, consistent with the Galaxy rotation curve.
Our sample contains clusters which are located much further away, and
at higher Galactic latitudes. We therefore expect to find
deviations from purely circular motions. Although these deviations
might be important signatures of accretion, one has to be very
carefully analysing distant star clusters in the anticenter, since 
both their
radial velocity is generally dominated by non circular motions and
the disk structure is more complicated than in the solar neighborhood.\\

Anyhow, we derive circular velocities for these clusters, using
Olling and Merrifield's (2000) equations 1 and 2, which combined together
yield:

\begin{equation}
V_{ROT} = \frac{R_{GC}}{8.5} \cdot [220 + \frac{RV}{\sin{l}\cos{b}}],
\end{equation}

\noindent
where the circular speed and distance from the Galactic Center of the
Sun are taken to be 220 km sec$^{-1}$ and 8.5 kpc, for consistency with the
adopted values to derive cluster Galactic coordinates (see
Sect.~5). Finally,
RV here represents the Local Standard of Rest (LSR) corrected radial velocity.\\

Figure 8 shows the derived rotational velocity distribution of the clusters as a function
of Galactocentric radius, compared to the Galaxy rotation curve from Olling and Merrifield (2000;
dashed line). The position of the Sun is indicated at (8.5,200), while
the range of values derived by Scott et al. (1995) for old open
clusters is represented by the thick horizontal segment.

All the clusters rotate slower than predicted by the rotation curve, apart from
Berkeley~29 and Berkeley 21, which have a significantly larger rotation
velocity, but lie very close to the Galactic anticenter.
Besides, the deviation from the Galaxy rotation curve increases with distance.

The mean rotational
velocity of our clusters is about 150 km/sec, which
differs from that
of old open clusters in the solar vicinity (Scott et al. 1995),
which rotate at the same speed as the Sun.

Before deriving conclusions on the meaning of this deviation,
it is worth keeping in mind that these clusters are on the average
much farther from the Galactic plane than the solar vicinity clusters,
and their kinematics is more complex, possibly due to the warping and flaring
of the disk.

To fully explore their kinematics we would
need proper motion measurements, to derive cluster orbits and
eccentricities (Carraro \& Chiosi 1994; Frinchaboy 2006; Casetti-Dinescu et al. 2007).
These quantities are mandatory to address the issue of the origin
of these clusters, and hopefully disentangle between an extra-galactic origin,
or the kinematic influence by a flared/warped disk.\\

\noindent
At odd with the results from the chemical analysis, the study
of kinematics presented here cannot help much to put more
constraints on the formation, origin and evolution of the outer
disk old open clusters system.

\section{Kinematic association to MRi}
The Galactic Anticenter Stellar Structure (GASS), or MRi,
exhibits quite a tight relation between the velocity
corrected to the Galactic standard of rest and the longitude
in a sample of putative M giant members (Crane et al. 2003).
Based  on that, Frinchaboy et al. (2004, 2006) looked for star clusters
associated with this structure, and actually found that seven clusters
(both open and globulars) are consistent with GASS membership in the Galactic standard of
rest velocity (v$_{GSR}$) versus longitude plane.

Meanwhile, new  radial velocities have been acquired for some of these
clusters and others which are located in the MRi region.
This  permit us to revise the possible kinematic association of
old open clusters with GASS.

In fact, Frinchaboy et al. 2006 could provide evidence that Saurer~1
and Berkeley~29 are probably associated with GASS.

We can here extend that analysis to  the larger anti-center cluster
sample we have described in previous Sections.
To this aim, we computed v$_{GSR}$ of each
one of the five program clusters (see Table~7), and for the additional 8
clusters (Table~8).

The velocity  V$_{GSR}$ has been computed as:

\begin{equation}
 V_{GSR} = RV_H + 9\cdot cos(b)cos(l) + (220+12)\cdot
  cos(b)sin(l) +7\cdot sin(b)
\end{equation}

where (9,12,7) are the components of the Sun  peculiar motion 
- consistent with M giants from Majewski et
al. (2003)-, and 220 is the Local Standard of Rest
rotational velocity (see also Frinchaboy et al 2006, eq. 4). 

   \begin{figure*}
   \centering
   \includegraphics[width=18cm]{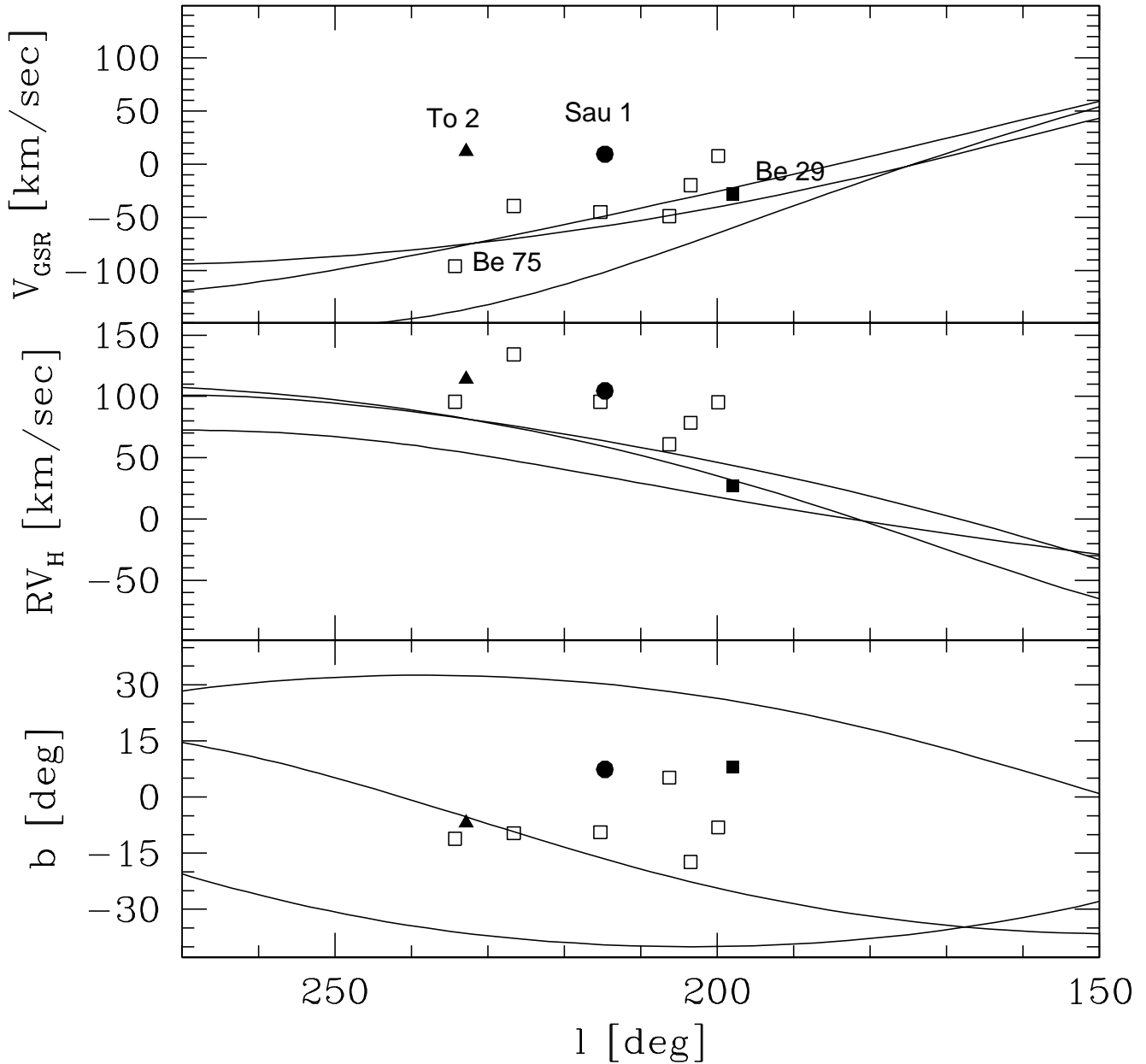}
      \caption{ Location and kinematics of old open clusters in the
      Galactic anti-center. Superimposed are Pe\~narubia et al. 2005
      models of the Monoceros Ring.
      The filled circle indicates Saurer~1, whilst the filled square
      Berkeley~29 and the filled triangle Tombaugh 2, previously
      indicated as MRi members.}
         \label{Figgrad}
   \end{figure*}

The aim is to compare cluster position and kinematics with the
expected trends for the MRi, following the same approach as in
Frinchaboy et al. (2006).
Theoretical --- $N$-body and analytical ---
calculations by
Pe\~narrubia et al. (2005) are available that predict the location and
kinematics of the MRi as a function of the shape of the Galactic halo.

Figure~9 compares the location and kinematics of old anti-center open
clusters having heliocentric distances larger than 7 kpc 
(small open squares), Berkeley 29, Tombaugh 2 and Saurer 1( MRi
candidates,
filled square, triangle  and circle, respectively) to the 
MRi models.
This cluster sample, 
the largest so far of distant anti-center star clusters -including Berkeley~29 and Saurer~1-
does not generally follow the expected MRi spatial distribution (bottom panel), 
while the heliocentric radial velocities (middle
panel) are only marginally compatible with MRi models, as are the Galactic standard of
rest velocities v$_{GSR}$ (top panel). \\
Finally, we notice that while Berkeley~29 seems to follow the MRi
models kinematics, but not the spatial location, 
Saurer~1 clearly deviates in all the diagrams, and Tombaugh~2 is only
spatially compatible with MRi  (but see Frinchaboy et al. 2007).
Berkeley~75  at l = $234^{o}$ seems to be the cluster that simultaneously 
best matches the model expectations in the three planes.

\noindent

This leads us to our final argument, now based on kinematics,
indicating that the preferred origin of our clusters is Galactic as
opposed to extra-galactic.

\section{Conclusions}
We have presented new radial velocities and metallicities derived from
high resolution spectra for five
old distant anti-center open clusters previously unstudied.
These results  have been combined with similar quality data
taken from the literature for eight additional old open clusters.
We discuss this material in the context of the chemical and kinematic
properties of the Galactic disk, with the principal aim
to understand whether these clusters are genuine disk
members, or that some of the clusters trace 
an extragalactic population which merged
into the disk a few Gyr ago, the MRi.

Our results can be summarized as follows.
\begin{itemize}
\item the spatial location of these clusters is compatible
with a warped/flared Galactic disk and poorly matches MRi model
predictions;
\item metallicities are compatible both with the Galactic disk and with
the M giants which trace MRi, whilst the abundance ratios of $\alpha$
elements are more compatible with the Galactic disk population;
\item the AMR is in better agreement with the Galactic disk than with
  MRi, although this latter is at present poorly constrained;
\item we find that the cluster kinematics are difficult to interpret,
in the sense that clusters seem to obey  the Galactic differential
rotation, but with increasing deviations at increasing distances.
However, it is difficult with present data to disentangle
whether these deviations are indicating an extragalactic origin, or
are simply a consequence of the complicated dynamics in a
warped/flared disk. In fact, in this case the expected increase of  
the velocity vertical component would naturally produce deviations
from purely circular orbits;
\item Models of MRi do not reproduce satisfactorily and simultaneously
the position, radial
velocities and Galactic standard of
rest velocities v$_{GSR}$ of the outer disk old open clusters here considered.
\item the radial abundance gradients, especially those for the 
$\alpha-$ elements, are in good agreement with prediction based on
simple Galactic disk models (Cescutti et al. 2006). We confirm that in the outer disk the
radial iron abundance gradient is flatter than in the solar vicinity.

\end{itemize}

This work emphasizes the difficulty to study the periphery of the
Galactic disk.
A better description of the properties of the warped/flared disk,
and of the Monoceros Ring models are mandatory to improve
the comparison with the observations that we are accumulating. At the same time,
the proper motion components of these clusters might greatly help
to understand their kinematic properties, especially near the Galactic anti-center.
The derivations of their orbits and eccentricities would
constitute an important step ahead in our knowledge of the
outer Galactic disk.

\begin{acknowledgements}
GC acknowledges fruitful discussions with Tom Richtler and thanks
Gabriele Cescutti for calculating the slope of the radial abundance
gradient for several elements.
DG gratefully acknowledges support from the Chilean
{\sl Centro de Astrof\'\i sica} FONDAP No. 15010003.
GC and DG acknowledge support from the Padova and Concepci\'on
Universities exchange program.  PMF is supported by
an NSF Astronomy and Astrophysics Postdoctoral Fellowship under award
AST-0602221.
Finally, the comments of the anonymous referee have been greatly appreciated.
\end{acknowledgements}

\input{tabline.tex}

\end{document}

%% file: tabline.tex
\longtab{3}{
\begin{longtable}{lllllllll}
\caption{}\\
\hline\hline
$\lambda$ &El & EP & log(gf) & Be25-12 & Be73-18 & Be75-22 & Rup4-3 & Rup7-2 \\
\hline
\endfirsthead
\caption{continued.}\\
\hline\hline
$\lambda$ & El & EP & log(gf) & Be25-12 & Be73-18 & Be75-22 & Rup4-3 & Rup7-2 \\
\hline
\endhead
\hline
\endfoot
4801.031 & 24.0 & 3.120 & -0.130 & 75.1 & 78.7 & 54.0 & 85.3 & 65.3 \\ 
4810.537 & 30.0 & 4.080 & -0.170 & 60.6 & 75.6 & 80.6 & 99.6 & 74.5 \\ 
4812.352 & 24.1 & 3.860 & -1.800 &  -  &  -  & 53.0 & 71.3 & 51.2 \\ 
4813.479 & 27.0 & 3.200 & 0.180 &  -  & 68.9 & 93.7 & 70.5 & 48.0 \\ 
4820.414 & 22.0 & 1.500 & -0.440 & 98.7 & 106.8 & 71.0 & 106.6 & 67.9 \\ 
4827.457 & 23.0 & 0.040 & -1.480 &  -  &  -  &  -  & 74.2 &  -  \\ 
4831.651 & 23.0 & 0.020 & -1.380 & 90.0 & 74.1 &  -  & 57.0 &  -  \\ 
4832.431 & 23.0 & 0.000 & -1.500 & 104.5 &  -  &  -  &  -  &  -  \\ 
4836.857 & 24.0 & 3.100 & -1.140 & 71.6 &  -  &  -  &  -  &  -  \\ 
4848.252 & 24.1 & 3.860 & -1.080 &  -  &  -  & 80.4 & 102.0 & 84.1 \\ 
4854.873 & 39.1 & 0.990 & -0.380 & 74.4 & 87.9 & 78.2 & 100.5 & 88.2 \\ 
4864.738 & 23.0 & 0.020 & -0.960 & 97.0 & 99.2 & 121.5 & 83.0 & 52.7 \\ 
4865.618 & 22.1 & 1.120 & -2.750 & 94.6 & 86.9 &  -  & 95.4 & 76.6 \\ 
4867.874 & 27.0 & 3.100 & 0.410 & 107.7 & 117.7 & 63.4 & 98.3 &  -  \\ 
4874.014 & 22.1 & 3.090 & -0.940 &  -  & 46.1 &  -  & 56.1 & 52.2 \\ 
4875.492 & 23.0 & 0.040 & -0.810 & 95.0 & 102.2 & 84.7 & 90.6 & 64.7 \\ 
4883.690 & 39.1 & 1.080 & 0.070 & 85.5 & 104.2 & 74.0 & 102.2 & 109.9 \\ 
4900.124 & 39.1 & 1.030 & -0.090 &  -  &  -  &  -  &  -  & 175.1 \\ 
4904.420 & 28.0 & 3.540 & -0.190 &  -  & 115.4 & 81.5 & 110.9 & 91.2 \\ 
4911.199 & 22.1 & 3.120 & -0.490 &  -  & 67.7 & 63.3 & 87.9 & 73.6 \\ 
4913.978 & 28.0 & 3.740 & -0.630 & 67.0 & 76.3 &  -  & 73.0 & 61.2 \\ 
4935.834 & 28.0 & 3.940 & -0.380 & 79.6 & 77.8 & 45.3 & 75.4 & 59.3 \\ 
4936.341 & 24.0 & 3.110 & -0.220 & 60.2 & 71.1 & 51.7 & 66.5 & 46.3 \\ 
4953.212 & 28.0 & 3.740 & -0.640 & 110.8 & 100.7 & 89.4 & 82.5 & 58.7 \\ 
4954.809 & 24.0 & 3.120 & -0.140 &  -  &  -  &  -  &  -  & 73.6 \\ 
4964.933 & 24.0 & 0.940 & -2.530 & 111.5 & 93.6 & 78.3 &  -  &  -  \\ 
4981.740 & 22.0 & 0.850 & 0.400 & 163.7 & 152.6 & 159.6 & 160.7 & 86.2 \\ 
4997.100 & 22.0 & 0.000 & -2.120 & 95.8 & 98.1 & 65.8 &  -  &  -  \\ 
4999.510 & 22.0 & 0.830 & 0.250 & 167.4 & 162.2 & 135.9 & 169.0 & 140.2 \\ 
5003.747 & 28.0 & 1.680 & -3.130 & 88.3 & 59.6 & 44.4 & 61.4 &  -  \\ 
5004.894 & 25.0 & 2.920 & -1.650 &  -  & 36.9 &  -  &  -  &  -  \\ 
5005.171 & 22.1 & 1.570 & -2.680 &  -  & 56.4 & 37.6 & 46.7 & 47.1 \\ 
5009.655 & 22.0 & 0.020 & -2.260 & 117.1 & 83.8 & 68.8 & 77.7 &  -  \\ 
5010.943 & 28.0 & 3.630 & -0.900 &  -  & 64.2 & 52.5 & 69.4 & 52.3 \\ 
5016.168 & 22.0 & 0.850 & -0.570 & 122.5 & 101.7 & 90.3 & 100.9 & 76.2 \\ 
5022.874 & 22.0 & 0.830 & -0.430 & 107.5 & 119.6 & 106.7 & 114.8 & 94.0 \\ 
5039.964 & 22.0 & 0.020 & -1.230 & 144.1 & 135.8 & 107.3 & 123.8 & 94.4 \\ 
5048.853 & 28.0 & 3.850 & -0.390 & 124.3 & 107.6 &  -  & 102.0 & 70.2 \\ 
5062.104 & 22.0 & 2.160 & -0.440 &  -  & 45.2 &  -  & 36.5 &  -  \\ 
5087.426 & 39.1 & 1.080 & -0.170 & 65.8 & 74.0 & 59.8 & 92.9 & 82.5 \\ 
5094.418 & 28.0 & 3.830 & -1.120 &  -  & 41.5 &  -  & 42.5 & 30.5 \\ 
5105.545 & 29.0 & 1.380 & -1.020 & 159.4 & 136.8 & 127.2 & 139.6 & 105.1 \\ 
5112.279 & 40.1 & 1.660 & -0.590 &  -  &  -  &  -  & 47.2 &  -  \\ 
5113.447 & 22.0 & 1.440 & -0.780 &  -  & 71.6 & 57.3 & 61.3 &  -  \\ 
5145.468 & 22.0 & 1.460 & -0.570 & 96.2 & 68.5 & 71.8 & 69.8 & 44.7 \\ 
5147.482 & 22.0 & 0.000 & -2.010 & 106.8 & 85.4 &  -  & 87.8 & 61.8 \\ 
5152.190 & 22.0 & 0.020 & -2.020 & 94.2 & 92.8 & 88.5 &  -  &  -  \\ 
5157.984 & 28.0 & 3.510 & -1.720 &  -  &  -  &  -  & 27.2 &  -  \\ 
5173.750 & 22.0 & 0.000 & -1.370 & 95.5 & 90.8 & 115.9 & 91.9 & 102.9 \\ 
5185.908 & 22.1 & 1.890 & -1.500 & 93.9 & 102.9 & 86.0 & 109.2 & 101.0 \\ 
5197.170 & 28.0 & 3.900 & -1.140 &  -  & 50.2 &  -  & 58.0 &  -  \\ 
5200.185 & 24.0 & 3.380 & -0.530 &  -  &  -  & 51.9 &  -  &  -  \\ 
5200.415 & 39.1 & 0.990 & -0.570 & 97.3 & 91.1 & 71.0 & 99.9 &  -  \\ 
5210.392 & 22.0 & 0.050 & -0.880 & 158.6 & 142.8 & 116.2 & 137.1 & 104.8 \\ 
5212.691 & 27.0 & 3.510 & -0.110 &  -  &  -  &  -  & 33.8 &  -  \\ 
5214.616 & 24.0 & 3.320 & -0.660 &  -  &  -  &  -  & 48.5 &  -  \\ 
5218.209 & 29.0 & 3.800 & 0.300 & 85.5 & 63.9 & 61.7 & 78.3 &  -  \\ 
5219.706 & 22.0 & 0.020 & -2.260 & 96.2 & 83.0 & 57.5 & 75.7 & 47.3 \\ 
5237.325 & 24.1 & 4.070 & -1.170 &  -  & 75.0 & 67.4 & 92.2 & 72.2 \\ 
5238.969 & 24.0 & 2.710 & -1.300 &  -  &  -  &  -  & 35.9 &  -  \\ 
5239.823 & 21.1 & 1.450 & -0.700 & 92.4 & 112.2 & 65.9 & 107.0 & 93.1 \\ 
5247.574 & 24.0 & 0.960 & -1.630 & 133.7 & 118.6 & 80.7 & 114.0 &  -  \\ 
5260.390 & 20.0 & 2.520 & -1.720 & 68.7 & 49.5 &  -  & 52.1 &  -  \\ 
5261.708 & 20.0 & 2.520 & -0.580 & 113.8 & 123.4 & 112.3 & 126.5 & 104.4 \\ 
5272.003 & 24.0 & 3.450 & -0.420 &  -  &  -  &  -  & 40.3 &  -  \\ 
5274.236 & 58.1 & 1.040 & 0.260 &  -  & 33.3 &  -  & 44.0 &  -  \\ 
5282.443 & 22.0 & 1.050 & -1.300 & 57.5 & 48.4 & 51.4 & 44.5 & 24.7 \\ 
5295.784 & 22.0 & 1.050 & -1.630 & 70.5 & 45.1 &  -  & 40.8 &  -  \\ 
5296.702 & 24.0 & 0.980 & -1.240 & 154.5 & 138.8 & 113.6 & 140.1 & 109.7 \\ 
5299.984 & 22.0 & 1.050 & -1.360 &  -  & 41.6 &  -  & 33.2 &  -  \\ 
5300.751 & 24.0 & 0.980 & -2.130 & 109.3 & 96.5 & 75.0 & 97.6 &  -  \\ 
5304.185 & 24.0 & 3.460 & -0.690 & 38.9 &  -  &  -  &  -  &  -  \\ 
5305.866 & 24.1 & 3.830 & -2.080 &  -  & 36.7 &  -  & 50.2 & 46.4 \\ 
5308.429 & 24.1 & 4.070 & -1.850 &  -  &  -  &  -  & 53.7 & 48.2 \\ 
5310.697 & 24.1 & 4.070 & -2.280 &  -  &  -  &  -  & 29.1 &  -  \\ 
5312.863 & 24.0 & 3.450 & -0.560 &  -  &  -  &  -  & 36.0 & 24.2 \\ 
5313.585 & 24.1 & 4.070 & -1.640 & 44.9 &  -  &  -  & 62.1 & 61.4 \\ 
5318.361 & 21.1 & 1.360 & -1.750 &  -  &  -  &  -  & 46.5 & 43.0 \\ 
5319.820 & 60.1 & 0.550 & -0.560 &  -  &  -  & 51.3 &  -  & 79.1 \\ 
5336.790 & 22.1 & 1.580 & -1.590 & 112.3 & 100.1 & 86.3 & 132.7 & 121.6 \\ 
5342.708 & 27.0 & 4.020 & 0.520 &  -  & 51.7 & 41.0 & 48.4 & 31.7 \\ 
5345.807 & 24.0 & 1.000 & -0.970 & 199.0 & 166.8 & 152.2 & 168.3 & 137.2 \\ 
5348.326 & 24.0 & 1.000 & -1.290 & 170.6 & 141.5 & 123.0 & 134.4 & 114.6 \\ 
5349.469 & 20.0 & 2.710 & -0.460 & 105.1 & 112.6 & 87.6 & 120.4 &  -  \\ 
5352.049 & 27.0 & 3.560 & 0.060 &  -  & 50.2 &  -  & 47.9 &  -  \\ 
5402.783 & 39.1 & 1.840 & -0.640 &  -  &  -  &  -  & 44.8 & 49.3 \\ 
5409.799 & 24.0 & 1.030 & -0.710 & 178.7 & 184.2 & 164.2 & 184.9 & 142.0 \\ 
5418.780 & 22.1 & 1.580 & -2.110 &  -  & 81.0 & 61.9 & 93.3 & 84.2 \\ 
5432.548 & 25.0 & 0.000 & -3.800 & 153.8 & 127.0 & 117.4 & 111.7 & 85.9 \\ 
5490.159 & 22.0 & 1.460 & -0.930 & 71.5 & 64.0 &  -  & 57.0 &  -  \\ 
5494.474 & 26.0 & 4.070 & -1.960 &  -  & 44.6 &  -  & 46.4 &  -  \\ 
5502.092 & 24.1 & 4.170 & -1.960 &  -  &  -  &  -  & 36.2 & 40.6 \\ 
5503.904 & 22.0 & 2.580 & -0.190 & 48.6 &  -  &  -  &  -  &  -  \\ 
5522.454 & 26.0 & 4.210 & -1.470 & 64.1 & 57.2 &  -  & 60.3 & 47.8 \\ 
5525.135 & 26.1 & 3.270 & -4.040 &  -  &  -  &  -  & 32.2 &  -  \\ 
5526.821 & 21.1 & 1.770 & 0.190 & 101.3 & 116.6 & 87.9 & 123.4 & 110.7 \\ 
5534.848 & 26.1 & 3.240 & -2.750 &  -  & 67.0 & 42.5 & 102.4 & 101.4 \\ 
5539.291 & 26.0 & 3.640 & -2.590 &  -  & 35.7 &  -  & 35.6 &  -  \\ 
5547.000 & 26.0 & 4.220 & -1.800 & 57.1 & 44.0 &  -  & 54.2 & 31.6 \\ 
5560.220 & 26.0 & 4.430 & -1.100 & 66.5 & 66.8 & 51.8 & 70.9 & 49.9 \\ 
5568.862 & 26.0 & 3.630 & -2.910 &  -  &  -  &  -  & 25.9 &  -  \\ 
5578.729 & 28.0 & 1.680 & -2.570 & 114.5 & 89.8 & 106.0 & 102.1 & 72.9 \\ 
5587.581 & 26.0 & 4.140 & -1.700 &  -  & 51.5 & 31.0 &  -  &  -  \\ 
5587.868 & 28.0 & 1.930 & -2.390 & 95.8 & 81.4 & 58.3 & 97.1 &  -  \\ 
5589.366 & 28.0 & 3.900 & -1.150 &  -  & 42.4 & 25.6 &  -  &  -  \\ 
5590.126 & 20.0 & 2.510 & -0.770 & 101.2 & 111.4 & 92.4 & 118.8 & 95.2 \\ 
5593.746 & 28.0 & 3.900 & -0.780 &  -  & 74.7 &  -  & 54.4 & 44.7 \\ 
5608.976 & 26.0 & 4.210 & -2.310 &  -  &  -  &  -  & 21.7 & 16.8 \\ 
5611.357 & 26.0 & 3.630 & -2.930 & 34.4 &  -  &  -  &  -  &  -  \\ 
5618.642 & 26.0 & 4.210 & -1.340 & 67.1 & 70.2 & 62.1 & 65.6 & 50.8 \\ 
5619.609 & 26.0 & 4.390 & -1.490 & 54.7 & 59.7 &  -  & 45.4 & 33.5 \\ 
5627.642 & 23.0 & 1.080 & -0.370 & 106.5 & 74.5 & 55.6 & 69.8 & 57.8 \\ 
5635.831 & 26.0 & 4.260 & -1.590 &  -  & 43.0 &  -  & 49.5 & 35.0 \\ 
5636.705 & 26.0 & 3.640 & -2.530 &  -  & 29.6 & 26.8 & 35.5 &  -  \\ 
5640.989 & 21.1 & 1.500 & -0.860 & 86.0 & 87.3 & 62.8 & 97.7 & 81.0 \\ 
5643.087 & 28.0 & 4.160 & -1.250 &  -  &  -  &  -  & 28.2 &  -  \\ 
5645.618 & 14.0 & 4.930 & -2.140 &  -  & 56.1 &  -  &  -  &  -  \\ 
5647.241 & 27.0 & 2.280 & -1.560 & 68.6 & 33.1 &  -  & 37.6 &  -  \\ 
5649.996 & 26.0 & 5.100 & -0.800 & 52.5 & 46.5 &  -  & 57.8 &  -  \\ 
5651.477 & 26.0 & 4.470 & -1.790 &  -  &  -  &  -  & 36.5 & 20.9 \\ 
5652.327 & 26.0 & 4.260 & -1.770 &  -  & 46.3 &  -  & 48.3 & 30.6 \\ 
5657.880 & 21.1 & 1.510 & -0.290 & 125.5 & 105.5 & 98.3 & 115.1 & 107.3 \\ 
5661.354 & 26.0 & 4.280 & -1.830 & 45.5 & 54.6 &  -  & 40.6 & 28.1 \\ 
5662.159 & 22.0 & 2.320 & -0.110 &  -  & 49.2 &  -  &  -  &  -  \\ 
5665.563 & 14.0 & 4.920 & -2.040 & 44.5 &  -  & 46.2 & 59.2 & 51.5 \\ 
5667.153 & 21.1 & 1.500 & -1.110 & 62.7 & 71.2 & 57.8 & 78.3 &  -  \\ 
5669.040 & 21.1 & 1.500 & -1.000 & 72.4 & 78.3 &  -  & 92.5 & 92.2 \\ 
5670.858 & 23.0 & 1.080 & -0.420 & 92.3 & 69.7 &  -  & 48.3 & 28.7 \\ 
5671.826 & 21.0 & 1.450 & 0.560 &  -  &  -  & 37.7 & 35.5 &  -  \\ 
5680.244 & 26.0 & 4.190 & -2.290 &  -  &  -  &  -  & 26.1 &  -  \\ 
5682.647 & 11.0 & 2.100 & -0.870 &  -  & 129.3 & 110.2 & 131.8 & 109.4 \\ 
5684.198 & 21.1 & 1.510 & -0.920 & 79.0 &  -  & 95.7 &  -  &  -  \\ 
5688.217 & 11.0 & 2.100 & -0.570 & 129.6 & 135.3 & 128.2 & 146.3 &  -  \\ 
5690.433 & 14.0 & 4.930 & -1.870 & 61.8 & 61.5 &  -  & 63.4 & 55.0 \\ 
5698.530 & 23.0 & 1.060 & -0.110 & 128.1 &  -  & 82.9 &  -  &  -  \\ 
5701.108 & 14.0 & 4.930 & -2.050 &  -  & 45.7 &  -  & 44.8 &  -  \\ 
5701.557 & 26.0 & 2.560 & -2.160 & 122.0 & 123.6 & 103.1 & 112.4 & 88.6 \\ 
5702.328 & 24.0 & 3.450 & -0.480 & 53.8 & 53.7 &  -  & 40.9 & 30.9 \\ 
5703.587 & 23.0 & 1.050 & -0.210 & 117.8 & 76.5 & 61.5 & 60.7 & 40.6 \\ 
5708.405 & 14.0 & 4.950 & -1.470 & 80.9 &  -  &  -  &  -  &  -  \\ 
5711.095 & 12.0 & 4.340 & -1.710 & 124.9 & 116.4 & 135.8 & 117.9 & 87.8 \\ 
5717.841 & 26.0 & 4.280 & -0.980 & 91.4 & 77.1 & 65.0 & 89.4 &  -  \\ 
5727.057 & 23.0 & 1.080 & -0.010 & 102.6 & 85.7 & 81.6 & 81.7 &  -  \\ 
5731.772 & 26.0 & 4.260 & -1.000 &  -  & 68.8 & 78.3 & 79.4 & 64.4 \\ 
5741.856 & 26.0 & 4.260 & -1.690 &  -  & 51.5 &  -  & 42.1 & 39.2 \\ 
5748.361 & 28.0 & 1.680 & -3.250 &  -  &  -  &  -  &  -  & 52.1 \\ 
5752.042 & 26.0 & 4.550 & -0.920 & 33.1 &  -  & 37.5 &  -  & 49.4 \\ 
5760.359 & 26.0 & 3.640 & -2.460 &  -  &  -  &  -  &  -  & 30.4 \\ 
5772.149 & 14.0 & 5.080 & -1.750 &  -  &  -  & 27.3 &  -  & 52.3 \\ 
5847.006 & 28.0 & 1.680 & -3.440 & 81.7 & 65.8 &  -  & 50.7 & 40.5 \\ 
5852.228 & 26.0 & 4.550 & -1.160 &  -  & 66.0 &  -  & 63.0 & 50.2 \\ 
5853.150 & 26.0 & 1.480 & -5.090 & 49.5 & 39.6 &  -  &  -  &  -  \\ 
5853.688 & 56.1 & 0.600 & -1.000 & 112.6 & 112.4 & 120.4 & 143.1 & 150.5 \\ 
5855.086 & 26.0 & 4.610 & -1.560 & 52.0 &  -  & 35.8 & 40.6 & 24.0 \\ 
5856.096 & 26.0 & 4.290 & -1.570 &  -  & 52.1 & 57.7 & 58.4 & 36.0 \\ 
5858.785 & 26.0 & 4.220 & -2.190 &  -  &  -  &  -  & 20.3 &  -  \\ 
5859.596 & 26.0 & 4.550 & -0.630 & 86.6 & 75.6 & 63.5 & 91.2 & 71.9 \\ 
5861.110 & 26.0 & 4.280 & -2.350 &  -  &  -  &  -  & 19.7 & 13.1 \\ 
5862.368 & 26.0 & 4.550 & -0.420 & 107.0 & 101.3 & 86.5 & 101.0 & 85.9 \\ 
5866.461 & 22.0 & 1.070 & -0.840 & 115.9 & 105.1 & 99.6 & 88.9 & 65.7 \\ 
5867.572 & 20.0 & 2.930 & -1.490 &  -  & 34.6 &  -  & 35.6 & 25.5 \\ 
5881.279 & 26.0 & 4.610 & -1.760 & 33.6 &  -  &  -  & 25.1 &  -  \\ 
5902.476 & 26.0 & 4.590 & -1.860 &  -  &  -  &  -  & 20.0 &  -  \\ 
5905.680 & 26.0 & 4.650 & -0.780 &  -  & 59.7 & 81.4 & 76.7 & 62.8 \\ 
5922.123 & 22.0 & 1.050 & -1.470 & 100.7 & 71.6 & 73.3 & 52.9 &  -  \\ 
5927.797 & 26.0 & 4.650 & -1.070 & 91.4 & 52.0 & 60.8 & 61.6 & 48.6 \\ 
5929.682 & 26.0 & 4.550 & -1.160 &  -  & 48.7 &  -  & 54.1 & 46.3 \\ 
5930.191 & 26.0 & 4.650 & -0.340 & 126.0 & 93.0 & 84.3 & 102.2 & 84.9 \\ 
5934.665 & 26.0 & 3.930 & -1.080 & 100.0 & 91.5 & 78.1 & 101.9 & 86.7 \\ 
5947.531 & 26.0 & 4.610 & -2.040 &  -  &  -  &  -  & 16.9 &  -  \\ 
5948.548 & 14.0 & 5.080 & -1.230 &  -  & 92.4 & 67.5 & 110.9 & 88.0 \\ 
5956.706 & 26.0 & 0.860 & -4.560 & 100.0 & 113.0 & 62.9 & 98.9 & 87.4 \\ 
5978.549 & 22.0 & 1.870 & -0.580 &  -  & 50.1 &  -  & 51.0 & 33.5 \\ 
5984.826 & 26.0 & 4.730 & -0.290 &  -  & 79.6 & 89.8 & 90.2 & 82.8 \\ 
5991.378 & 26.1 & 3.150 & -3.550 & 45.7 & 42.4 & 39.8 & 60.8 & 60.3 \\ 
5996.740 & 28.0 & 4.230 & -1.060 &  -  & 36.1 & 39.2 & 29.5 & 20.4 \\ 
6003.022 & 26.0 & 3.880 & -1.020 & 93.0 & 99.3 & 111.9 & 105.3 & 89.5 \\ 
6007.968 & 26.0 & 4.650 & -0.760 & 73.5 & 70.8 &  -  & 76.0 & 61.6 \\ 
6008.566 & 26.0 & 3.880 & -0.920 & 126.6 & 104.2 & 92.1 & 98.4 & 92.2 \\ 
6013.497 & 25.0 & 3.070 & -0.250 & 120.2 & 117.7 & 94.9 & 112.4 & 82.8 \\ 
6016.647 & 25.0 & 3.070 & -0.090 &  -  & 129.2 & 105.2 & 113.6 & 107.6 \\ 
6021.803 & 25.0 & 3.080 & 0.030 & 151.4 & 131.1 & 120.6 & 119.2 & 98.6 \\ 
6027.059 & 26.0 & 4.070 & -1.200 & 104.2 & 83.2 & 71.5 & 100.3 & 76.7 \\ 
6053.693 & 28.0 & 4.230 & -1.070 &  -  &  -  &  -  & 27.5 &  -  \\ 
6056.013 & 26.0 & 4.730 & -0.460 & 82.9 & 81.1 & 70.0 & 87.4 & 68.1 \\ 
6065.494 & 26.0 & 2.610 & -1.590 & 148.3 & 157.2 & 139.1 & 163.6 &  -  \\ 
6078.499 & 26.0 & 4.790 & -0.380 &  -  & 77.7 & 67.9 & 97.6 & 80.7 \\ 
6079.016 & 26.0 & 4.650 & -0.970 &  -  &  -  & 80.7 & 65.0 & 41.3 \\ 
6081.448 & 23.0 & 1.050 & -0.580 &  -  &  -  & 74.2 &  -  & 32.2 \\ 
6082.718 & 26.0 & 2.220 & -3.530 & 82.1 & 78.2 & 61.9 & 73.7 & 59.5 \\ 
6084.105 & 26.1 & 3.200 & -3.800 &  -  &  -  &  -  & 50.0 & 61.8 \\ 
6086.288 & 28.0 & 4.260 & -0.470 &  -  & 50.4 &  -  & 53.9 & 47.5 \\ 
6089.574 & 26.0 & 5.020 & -0.870 & 58.9 & 53.8 &  -  & 66.8 & 47.2 \\ 
6090.216 & 23.0 & 1.080 & -0.060 & 85.4 & 85.1 & 57.8 & 68.9 & 59.9 \\ 
6091.177 & 22.0 & 2.270 & -0.420 &  -  & 36.5 &  -  & 38.5 & 19.9 \\ 
6093.649 & 26.0 & 4.610 & -1.320 &  -  & 42.4 &  -  & 48.7 & 36.7 \\ 
6094.377 & 26.0 & 4.650 & -1.560 &  -  & 32.2 &  -  & 36.1 & 18.9 \\ 
6096.671 & 26.0 & 3.980 & -1.760 & 58.0 & 63.5 & 72.4 & 61.2 & 44.6 \\ 
6098.250 & 26.0 & 4.560 & -1.810 &  -  &  -  & 27.4 & 38.9 &  -  \\ 
6108.125 & 28.0 & 1.680 & -2.470 & 98.8 & 102.5 & 89.5 & 107.1 & 85.1 \\ 
6111.078 & 28.0 & 4.090 & -0.830 &  -  & 51.9 & 45.6 & 50.9 & 31.7 \\ 
6113.329 & 26.1 & 3.210 & -4.120 &  -  &  -  &  -  & 37.4 &  -  \\ 
6119.532 & 23.0 & 1.060 & -0.320 & 99.6 & 59.4 &  -  & 58.6 & 42.1 \\ 
6120.258 & 26.0 & 0.910 & -5.860 & 57.2 & 28.0 &  -  & 27.5 &  -  \\ 
6125.026 & 14.0 & 5.610 & -1.570 & 49.2 &  -  &  -  & 39.4 & 34.9 \\ 
6126.224 & 22.0 & 1.070 & -1.420 & 94.7 & 73.2 & 57.9 & 57.9 & 44.5 \\ 
6128.984 & 28.0 & 1.680 & -3.390 & 85.4 & 74.5 &  -  & 62.7 & 49.0 \\ 
6130.141 & 28.0 & 4.260 & -0.980 &  -  & 43.5 &  -  &  -  & 25.5 \\ 
6137.002 & 26.0 & 2.200 & -2.910 & 106.7 & 104.9 & 99.7 &  -  &  -  \\ 
6145.020 & 14.0 & 5.610 & -1.440 &  -  &  -  &  -  & 45.8 &  -  \\ 
6149.249 & 26.1 & 3.890 & -2.720 & 53.4 & 39.2 & 37.7 & 61.4 & 69.8 \\ 
6151.623 & 26.0 & 2.180 & -3.260 & 114.9 & 83.8 & 83.1 & 95.5 & 71.7 \\ 
6154.230 & 11.0 & 2.100 & -1.570 & 67.8 & 51.7 & 60.6 & 62.9 & 46.1 \\ 
6157.733 & 26.0 & 4.070 & -1.260 &  -  & 96.1 & 77.2 & 91.9 & 80.7 \\ 
6160.753 & 11.0 & 2.100 & -1.260 & 69.7 & 80.7 & 55.7 & 80.7 & 67.3 \\ 
6161.295 & 20.0 & 2.520 & -1.270 & 76.0 & 79.8 & 75.4 & 87.0 & 80.1 \\ 
6165.363 & 26.0 & 4.140 & -1.480 & 66.4 & 64.3 & 61.3 & 59.8 & 54.2 \\ 
6166.440 & 20.0 & 2.520 & -1.140 & 97.5 & 87.7 & 83.3 & 91.2 & 72.2 \\ 
6169.044 & 20.0 & 2.520 & -0.800 & 136.6 & 93.2 & 103.3 & 112.0 & 100.7 \\ 
6169.564 & 20.0 & 2.520 & -0.480 & 151.9 & 121.0 & 123.3 & 123.1 & 105.7 \\ 
6173.341 & 26.0 & 2.220 & -2.840 & 107.5 & 104.1 & 79.5 & 107.7 & 92.9 \\ 
6176.816 & 28.0 & 4.090 & -0.240 & 69.0 & 79.6 & 70.5 & 76.7 & 62.7 \\ 
6177.253 & 28.0 & 1.830 & -3.600 & 52.2 & 41.5 &  -  & 33.7 &  -  \\ 
6186.717 & 28.0 & 4.100 & -0.900 &  -  &  -  &  -  & 43.7 & 34.0 \\ 
6187.995 & 26.0 & 3.940 & -1.600 & 76.3 & 66.8 &  -  & 72.1 & 55.2 \\ 
6199.186 & 23.0 & 0.290 & -1.280 &  -  & 73.6 &  -  & 41.8 & 33.5 \\ 
6200.321 & 26.0 & 2.610 & -2.390 &  -  & 111.0 &  -  & 114.3 & 97.2 \\ 
6204.610 & 28.0 & 4.090 & -1.150 &  -  &  -  &  -  & 40.9 & 29.3 \\ 
6213.437 & 26.0 & 2.220 & -2.540 & 130.8 & 131.2 & 94.5 & 122.4 & 109.0 \\ 
6216.358 & 23.0 & 0.280 & -0.810 & 109.3 & 82.7 &  -  & 66.9 & 46.4 \\ 
6219.287 & 26.0 & 2.200 & -2.390 & 113.5 & 125.6 & 102.3 & 135.1 & 120.6 \\ 
6223.990 & 28.0 & 4.100 & -0.970 &  -  & 37.9 &  -  & 40.3 & 28.9 \\ 
6226.740 & 26.0 & 3.880 & -2.080 & 43.3 & 44.1 &  -  & 40.2 & 33.1 \\ 
6230.098 & 28.0 & 4.100 & -1.200 &  -  &  -  &  -  &  -  & 26.0 \\ 
6232.648 & 26.0 & 3.650 & -1.060 & 113.4 & 79.0 & 100.0 & 113.0 & 96.0 \\ 
6240.653 & 26.0 & 2.220 & -3.230 & 99.3 & 88.5 & 77.2 & 90.1 & 81.9 \\ 
6243.114 & 23.0 & 0.300 & -0.980 & 152.2 & 97.0 & 85.4 & 82.1 &  -  \\ 
6245.620 & 21.1 & 1.510 & -1.050 & 61.6 & 75.9 & 55.6 & 76.9 & 82.6 \\ 
6246.327 & 26.0 & 3.600 & -0.730 & 154.9 & 123.2 & 113.4 & 134.7 & 119.2 \\ 
6247.562 & 26.1 & 3.870 & -2.320 & 57.9 & 60.8 &  -  & 65.7 & 92.7 \\ 
6251.825 & 23.0 & 0.290 & -1.340 & 112.2 & 71.4 & 43.0 & 51.7 & 29.3 \\ 
6252.565 & 26.0 & 2.400 & -1.740 & 159.7 & 175.2 & 122.9 & 166.4 & 142.5 \\ 
6258.110 & 22.0 & 1.440 & -0.360 & 98.9 & 94.1 & 76.0 & 86.2 & 65.4 \\ 
6261.106 & 22.0 & 1.430 & -0.480 & 145.5 & 105.0 & 82.2 & 94.7 & 74.0 \\ 
6265.141 & 26.0 & 2.180 & -2.510 & 161.3 & 133.5 & 99.9 & 133.5 & 124.5 \\ 
6270.231 & 26.0 & 2.860 & -2.550 & 108.8 & 91.8 & 79.0 & 92.3 & 77.1 \\ 
6279.740 & 21.1 & 1.500 & -1.160 & 55.3 &  -  &  -  & 78.6 &  -  \\ 
6280.622 & 26.0 & 0.860 & -4.340 & 110.6 & 114.0 & 93.0 & 135.7 & 101.7 \\ 
6297.799 & 26.0 & 2.220 & -2.700 &  -  & 127.3 & 113.7 & 125.3 & 130.5 \\ 
6300.311 & 8.0 & 0.000 & -9.750 &  -  &  -  &  -  & 23.2 &  -  \\ 
6301.508 & 26.0 & 3.650 & -0.570 & 126.6 & 120.3 & 124.6 & 146.4 & 129.0 \\ 
6311.504 & 26.0 & 2.830 & -3.160 & 104.1 & 58.0 & 44.4 & 62.0 & 47.7 \\ 
6315.814 & 26.0 & 4.070 & -1.670 & 74.1 & 64.2 & 47.9 & 66.9 & 50.1 \\ 
6318.708 & 12.0 & 5.110 & -1.970 &  -  &  -  &  -  &  -  & 38.1 \\ 
6322.694 & 26.0 & 2.590 & -2.380 & 132.8 & 109.6 & 97.3 & 109.0 & 96.0 \\ 
6327.604 & 28.0 & 1.680 & -3.080 & 68.5 & 91.4 & 61.2 & 76.2 & 71.1 \\ 
6330.096 & 24.0 & 0.940 & -2.870 &  -  & 71.9 &  -  & 72.0 & 54.8 \\ 
6330.852 & 26.0 & 4.730 & -1.220 &  -  & 54.3 & 71.9 & 49.7 & 43.9 \\ 
6335.337 & 26.0 & 2.200 & -2.280 & 145.9 & 137.8 & 117.1 & 135.9 & 126.6 \\ 
6369.463 & 26.1 & 2.890 & -4.210 &  -  & 25.6 & 37.6 & 44.7 & 37.4 \\ 
6378.256 & 28.0 & 4.150 & -0.820 &  -  & 43.7 & 49.0 & 46.1 &  -  \\ 
6380.750 & 26.0 & 4.190 & -1.340 & 88.6 & 71.7 & 54.4 & 81.2 & 63.9 \\ 
6383.715 & 26.1 & 5.550 & -2.090 &  -  &  -  &  -  &  -  & 15.4 \\ 
6384.668 & 28.0 & 4.150 & -1.000 &  -  &  -  &  -  & 38.0 & 28.8 \\ 
6392.538 & 26.0 & 2.280 & -3.970 &  -  &  -  &  -  & 45.4 &  -  \\ 
6393.612 & 26.0 & 2.430 & -1.630 & 176.7 & 192.0 & 131.1 & 171.0 & 151.5 \\ 
6411.658 & 26.0 & 3.650 & -0.700 & 146.6 & 125.8 & 111.1 & 139.9 & 125.7 \\ 
6416.928 & 26.1 & 3.890 & -2.700 &  -  & 47.0 &  -  & 61.9 & 64.1 \\ 
6432.683 & 26.1 & 2.890 & -3.580 & 41.0 & 53.3 & 57.0 & 73.6 & 76.6 \\ 
6436.411 & 26.0 & 4.190 & -2.400 &  -  & 30.6 &  -  & 25.9 & 15.1 \\ 
6439.083 & 20.0 & 2.520 & 0.190 & 159.8 & 172.7 & 158.2 & 189.5 & 162.8 \\ 
6455.605 & 20.0 & 2.520 & -1.290 & 119.2 & 75.0 & 45.3 & 89.7 & 66.7 \\ 
6456.391 & 26.1 & 3.900 & -2.100 & 72.7 & 67.0 &  -  & 88.5 & 104.3 \\ 
6471.668 & 20.0 & 2.520 & -0.690 & 102.3 & 110.0 & 108.8 & 131.6 & 111.1 \\ 
6481.878 & 26.0 & 2.280 & -2.940 & 113.4 & 95.1 & 96.5 & 108.5 & 101.3 \\ 
6482.809 & 28.0 & 1.930 & -2.780 & 79.4 & 78.7 &  -  & 79.7 & 68.9 \\ 
6493.788 & 20.0 & 2.520 & -0.110 & 143.4 & 134.0 & 133.3 & 146.2 & 132.7 \\ 
6496.908 & 56.1 & 0.600 & -0.380 & 153.5 & 141.7 & 142.2 & 200.6 &  -  \\ 
6498.945 & 26.0 & 0.960 & -4.660 & 112.4 & 87.1 & 97.1 & 101.3 & 84.5 \\ 
6499.654 & 20.0 & 2.520 & -0.820 & 103.6 & 114.1 & 105.0 & 121.0 & 96.9 \\ 
6516.083 & 26.1 & 2.890 & -3.380 & 68.0 & 71.6 &  -  & 104.8 & 95.2 \\ 
6518.373 & 26.0 & 2.830 & -2.560 & 128.8 & 77.7 & 81.6 & 104.2 &  -  \\ 
6533.940 & 26.0 & 4.560 & -1.200 &  -  & 57.0 &  -  & 62.8 &  -  \\ 
6554.238 & 22.0 & 1.440 & -1.220 & 88.5 & 51.7 &  -  & 41.8 &  -  \\ 
6572.795 & 20.0 & 0.000 & -4.320 & 135.9 & 94.1 & 89.8 & 79.1 & 57.0 \\ 
6574.254 & 26.0 & 0.990 & -4.960 & 123.9 & 83.4 & 91.6 & 85.3 & 60.0 \\ 
6581.218 & 26.0 & 1.480 & -4.680 &  -  & 65.3 & 51.2 & 60.4 & 52.9 \\ 
6586.319 & 28.0 & 1.950 & -2.780 & 80.6 & 64.7 & 66.8 & 76.6 & 59.2 \\ 
6593.884 & 26.0 & 2.430 & -2.300 & 118.0 & 136.0 &  -  & 128.1 & 113.7 \\ 
6598.611 & 28.0 & 4.230 & -0.930 &  -  &  -  &  -  & 38.5 &  -  \\ 
6604.600 & 21.1 & 1.360 & -1.140 &  -  & 73.6 &  -  & 85.1 & 86.3 \\ 
6606.979 & 22.1 & 2.060 & -2.900 &  -  &  -  &  -  & 35.5 &  -  \\ 
6608.044 & 26.0 & 2.280 & -3.960 &  -  & 52.4 &  -  & 47.0 & 30.6 \\ 
6609.118 & 26.0 & 2.560 & -2.650 &  -  & 98.7 & 105.1 & 105.8 & 92.0 \\ 
6625.039 & 26.0 & 1.010 & -5.320 & 90.6 & 55.3 & 46.8 & 57.9 & 39.7 \\ 
6627.560 & 26.0 & 4.550 & -1.500 &  -  & 36.0 & 44.0 & 50.0 & 33.9 \\ 
6633.758 & 26.0 & 4.560 & -0.810 & 63.7 & 82.4 & 65.4 &  -  &  -  \\ 
6635.137 & 28.0 & 4.420 & -0.750 &  -  & 39.7 &  -  & 41.2 & 21.1 \\ 
6696.032 & 13.0 & 3.140 & -1.320 &  -  & 58.0 &  -  & 57.6 & 50.8 \\ 
6698.669 & 13.0 & 3.140 & -1.620 &  -  & 28.6 &  -  &  -  &  -  \\ 
6703.576 & 26.0 & 2.760 & -3.000 & 87.9 & 69.9 & 47.8 & 77.8 & 55.2 \\ 
6713.745 & 26.0 & 4.790 & -1.410 &  -  & 34.7 &  -  & 36.1 & 22.9 \\ 
6725.364 & 26.0 & 4.100 & -2.210 & 48.5 & 27.1 &  -  & 33.0 & 23.5 \\ 
6726.673 & 26.0 & 4.610 & -1.050 & 41.8 & 67.5 & 45.6 & 59.8 & 53.1 \\ 
6733.153 & 26.0 & 4.640 & -1.440 &  -  &  -  &  -  & 44.0 & 31.0 \\ 
6739.524 & 26.0 & 1.560 & -4.850 & 72.8 & 48.5 &  -  & 37.2 & 25.4 \\ 
6750.164 & 26.0 & 2.420 & -2.580 &  -  & 112.2 & 93.9 & 126.3 & 104.7 \\ 
6767.784 & 28.0 & 1.830 & -2.060 & 124.8 & 119.0 & 83.7 & 119.0 & 105.8 \\ 
6772.321 & 28.0 & 3.660 & -0.960 & 61.1 & 68.4 & 55.5 & 74.9 & 54.7 \\ 
6786.860 & 26.0 & 4.190 & -1.900 & 62.2 & 39.0 &  -  & 40.7 & 29.5 \\ 
\end{longtable}
}